\documentclass[aps,prd,twocolumn,superscriptaddress,nofootinbib]{revtex4-2}
\usepackage{graphicx}
\usepackage{amsmath}
\usepackage{amssymb}
\usepackage{xcolor}
\usepackage{placeins}
\usepackage[hidelinks]{hyperref}

\usepackage[normalem]{ulem}
\usepackage{tikz}
\usetikzlibrary{arrows.meta,decorations.markings}
\tikzset{leg/.style={line width=0.5pt},
  arw/.style={postaction={decorate,decoration={markings,mark=at position #1 with {\arrow{Latex[length=1.7mm]}}}}}}

\newcommand{\github}{\raisebox{-0.35ex}{\includegraphics[height=2.0ex]{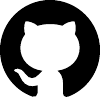}}}

\graphicspath{{figures_v2/}}

\newcommand{\Mc}{\mathcal{M}_c}

\begin{document}

\title{Unifying stochastic and deterministic signals: a practical search for the\\ intermittent binary black-hole background}

\author{T\"{o}re Boybeyi}
\affiliation{School of Physics and Astronomy, University of
Minnesota, Minneapolis, Minnesota 55455, USA}
\author{Vuk Mandic}
\affiliation{School of Physics and Astronomy, University of
Minnesota, Minneapolis, Minnesota 55455, USA}
\author{Joseph D. Romano}
\affiliation{Department of Physics and Astronomy, University of
Texas Rio Grande Valley, One West University Boulevard,
Brownsville, TX 78520, USA}

\date{\today}

\begin{abstract}
Over the past decade, the Advanced LIGO, Advanced Virgo, and KAGRA detectors have detected nearly 400 compact binary mergers, most of which are stellar-mass binary black-hole (BBH) mergers. Nevertheless, most mergers in the Universe are too weak to be individually detected at current sensitivity levels. Since a BBH merger in band is much shorter than the average time between successive mergers, the population of these subthreshold events gives rise to an intermittent background of gravitational waves (GWs). This GW background is guaranteed to exist, but it has not yet been detected with current stochastic or deterministic searches. Here we propose a fast and accurate likelihood-based search for this background, which unifies the standard model-agnostic cross-correlation search and a fully deterministic, template-based search in a single likelihood. Our search is based on a truncated likelihood ratio, which is fully marginalized over the extrinsic parameters of the individual mergers, and which represents the data in two detectors as an expansion in terms of approximate sufficient statistics. These statistics are higher-order cross-moments of the data, weighted by geometrical factors that generalize the standard overlap reduction function (ORF) used in cross-correlation searches. We show that almost all the Bayesian evidence and information on the BBH merger population comes from correlations between frequency bins induced by individual mergers and is captured in the higher-order generalized ORFs. Consequently, any likelihood function that does not take these inter-frequency correlations into account is necessarily suboptimal. We validate our search on one year of simulated Advanced LIGO data, and find that a catalog-informed analysis, which uses information from both the detected and unresolved mergers, can infer the peak of the redshift distribution of the population of BBH mergers.
\end{abstract}

\maketitle
\tableofcontents

\section{Introduction}

The gravitational-wave (GW) background from compact binary mergers is one target of current searches with the Advanced LIGO~\cite{LIGOScientific:2014pky}, Advanced Virgo~\cite{VIRGO:2014yos}, and KAGRA~\cite{KAGRA:2020tym} detectors~\cite{Regimbau:2008nj,Regimbau2011,MeacherThraneRegimbau,Meacher:2015iua,LVKO3iso,Virgo:2025aai,LIGOScientific:2026mjf,Mukherjee2020timedep,Sah2023nonstationary,Sah2025upperbound}.
At the heart of any such search is the likelihood of the data in the presence of a signal, marginalized over the parameters of the sources.
There are two main ways to model these signals.
(i) In the fully stochastic approach, the signal is treated as a Gaussian random process~\cite{Lawrence2023}, and the data of two detectors then enter only through their cross-correlation, weighted by the overlap reduction function (ORF)~\cite{Allen1999,Romano2017}.
(ii) In the fully deterministic approach, the signal is a waveform with definite parameters, whose model enters the likelihood of every segment of data~\cite{Smith2018,Kou2025}.

In this paper we ask how the description of the data in terms of its likelihood function passes from the fully stochastic approach to the fully deterministic approach as the signal becomes louder~\cite{Cornish:2013nma,Cornish:2015pda}.
For a weak signal, the likelihood surface is nearly flat and can be compressed into a few statistics of the data, which is why the cross-correlation works so well in this limit~\cite{Matas:2020roi}.
For a loud signal, the likelihood surface is sharply peaked, and only the full waveform resolves its peak.
Between these two limits, the phases of the waveform correlate the frequency bins with each other, and as the signal-to-noise ratio (SNR) grows the description has to keep more and more statistics that carry these correlations.

\begin{figure*}[!tbp]
\centering
\includegraphics[width=0.9\textwidth]{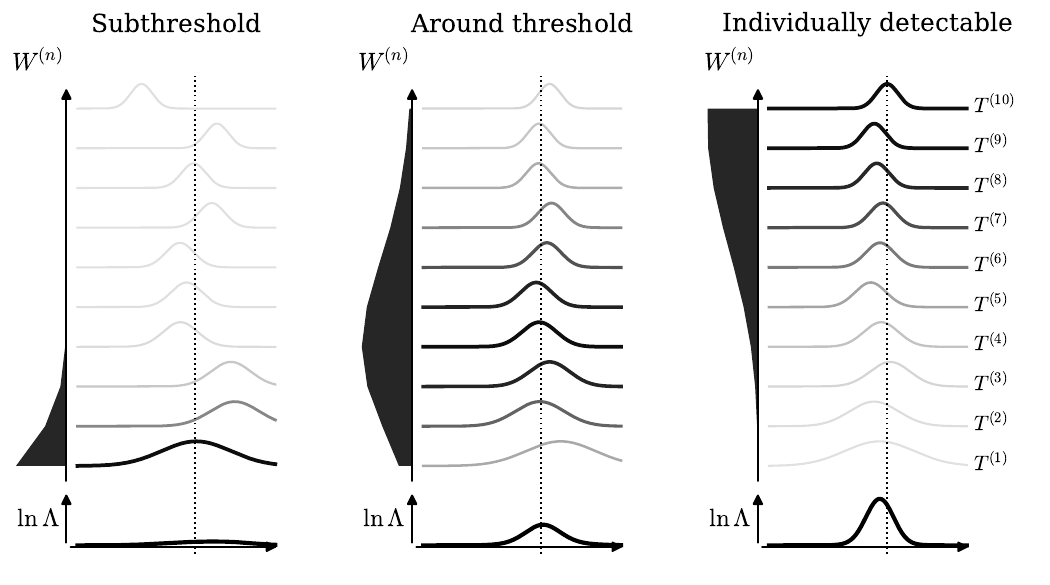}
\caption{Schematic of the likelihood ratio $\Lambda$ at three SNRs that increase from left to right, with a plot of $\ln\Lambda$ in the bottom row.
Each row is a summary statistic $T^{(n)}$, and its shade and the profile on the left give its weight $W^{(n)}$.
At low SNR, only the first-order statistic has a non-negligible weight, while the peaks of the higher-order statistics scatter with the noise.
At high SNR, the higher-order statistics dominate and peak together at the true value.}
\label{fig:schematic}
\end{figure*}

In the following sections, we answer the  above question by expanding the marginalized likelihood of a single merger in powers of the signal, and we formalize the procedure that finds the summary statistics of the data on which it depends. In a companion paper~\cite{companion}, we wrote the likelihood of the background in terms of the characteristic function of the strain of a single merger. Each approximation of this function gives a search with its own set of statistics. This paper builds one of these searches for a pair of detectors and applies it to a realistic population.
We find that these summary statistics, denoted $T^{(n)}_{ab}$, are cross moments of the data of the two detectors, delayed relative to one another, where $a$ and $b$ count the factors taken from the first detector.
The geometry of the detectors and the extrinsic parameters of the source (e.g., its sky position, orientation, and distance), enter through functions $W^{(n)}_{ab}$ of the delay, which can be interpreted as {\it generalized} overlap reduction functions (GORFs).
For example, at first order and for weak signals, the Fourier transform of $W^{(1)}_{10}$ reduces to the standard ORF in the literature.
In addition, we find that the likelihood ratio of a signal with matched-filter SNR $\rho_{\rm mf}$ saturates around the order $n \approx \rho^2_{\rm mf}/2$, illustrated schematically in Fig.~\ref{fig:schematic}.

These observations allow us to construct a single search in terms of these statistics, which keeps the computational advantage of the cross-correlation search while retaining the full information of the deterministic search for signals below an SNR that grows with the order of the series.
The marginalization over the sky position, the orientation and the distance is contained in the GORFs, which for a given detector pair depend only on the filter powers (see \eqref{eq:zu}) and the distance prior (or the distance itself when it is kept as a parameter).
We can therefore pre-compute the weights once before analyzing the data, so that the evaluation of a segment does not need to be sampled over the extrinsic parameters.
Fast marginalizations of the extrinsic parameters already exist for single events, using semi-analytic methods or importance sampling~\cite{Islam:2022afg,Roulet:2024hwz}, or using neural networks~\cite{Dax:2021tsq}.
But, the above methods are designed for {\it detected} events, whose likelihood has a clear peak, while most segments of a background search contain only subthreshold signals, if any.

We apply our proposed search to a year of simulated data for the two LIGO detectors
at O4 sensitivity~\cite{LIGOScientific:2014pky,Capote2025,Soni2025}, including an injected population of binary black hole mergers.
We show that our method detects the background and recovers the merger rate, and that the information it infers about the population saturates at a finite-order of the series.
We also show that this information comes from correlations between frequency bins, which the cross-correlation leaves out, and that about one third of the information comes from the mergers below the threshold of individual detections.
We find that a catalog-informed analysis, which uses information from both the detected and unresolved mergers, is capable of inferring the peak of the redshift distribution of the population of BBH mergers (Fig.~\ref{fig:pop}).
This is not possible if we analyze just the detected or the unresolved mergers separately.
On single events, our likelihood ratio agrees with nested sampling (Appendices~\ref{app:BF-convergence} and~\ref{app:ns-agreement}).

The rest of the paper is organized as follows.
In Sec.~\ref{sec:method}, we derive the marginalized likelihood ratio, showing how it can be written as an expansion in terms of a series of sufficient statistics $T^{(n)}_{ab}$ multiplied by corresponding weights $W^{(n)}_{ab}$.
In Sec.~\ref{sec:population}, we validate our proposed method on one year of simulated data, showing that we can infer the population parameters for the intermittent background of subthreshold events consistent with the injected values, and that including correlations between frequency bins is a crucial component for detecting the background.
We conclude with a brief summary in Sec.~\ref{sec:conclusion}.
In Appendix~\ref{app:formalism}, we
provide an alternative derivation of the likelihood ratio, which starts
from the characteristic function of the strain.
Appendix~\ref{app:norm} shows how to normalize the truncated likelihood ratio, and Appendix~\ref{app:construction} how we precompute the GORFs.
In Appendix~\ref{app:nstat}, we show how the search passes from the stochastic to the deterministic description of the data as the SNR of an event grows and compare it with nested sampling.
Appendix~\ref{app:pop-priors} gives the population model and its priors.
In Appendix~\ref{app:other-cases}, we repeat the analysis of Sec.~\ref{sec:population} for other truncation orders and detection thresholds, and in Appendix~\ref{app:null}, we apply the search to several years of Gaussian noise without mergers.

\section{Calculating the marginalized likelihood ratio}
\label{sec:method}

We divide the data from two detectors ($I=1,2$) into short segments, each containing either noise alone or noise together with a merger signal~\cite{Drasco2003}.
Since the mergers arrive at random, the number of mergers in a segment follows a Poisson distribution with mean $\bar N = R_0 V T_{\rm seg}$, where $R_0$ is the merger rate density, $V$ is the volume covered by the distance prior defined below, and $T_{\rm seg}$ is the segment duration.
(Note that $V = \int dz\, (dV_c/dz)\, \psi(z)/(1+z)$ for a rate $R_0\,\psi(z)$ that evolves with redshift $z$.)
The likelihood of the data is then
\begin{equation}
\mathcal{L}(\{d\} \mid \bar N) = \prod_i \Big[
e^{-\bar N}\, \mathcal{L}_n(d_i)
+ \bar N e^{-\bar N}\, \mathcal{L}_s(d_i)+\, \mathcal{O}(\bar N^2) \Big]\,.
\label{eq:mixture}
\end{equation}
For small $\bar N$ this is the familiar duty cycle model with $\xi \approx \bar N$~\cite{Drasco2003,Thrane:2013kb,Smith2018}, although $\xi$ and $\bar N$ differ in general.
We use $\bar N$ because it is tied directly to the merger rate.
For binary black holes and segments of a few seconds, $\bar N \ll 1$.
So we neglect segments with two or more mergers\footnote{These segments enter with the Poisson probability $\bar N^2 e^{-\bar N}/2 \simeq 2 \times 10^{-4}$ for the $\bar N = 0.019$ of the one year of simulated data described in Sec.~\ref{sec:population}.},
for which $\mathcal{L}(\{d\})$ in \eqref{eq:mixture} reduces to
\begin{equation}
\mathcal{L}(\{d\} \mid \bar N) = \prod_i \Big[
(1-\bar N)\, \mathcal{L}_n(d_i)
+ \bar N \, \mathcal{L}_s(d_i)\big]\,.
\label{eq:mixture2}
\end{equation}
This mixture is the first order in $\bar N$ of the likelihood of Ref.~\cite{companion}. The signal likelihood of one segment $d \equiv d_i$ is then an integral over the parameters of the merger,
\begin{equation}
\mathcal{L}_s(d) = \int d\vec\lambda_{\rm in}\; \pi(\vec\lambda_{\rm in})
\int d\vec\lambda_{\rm ex}\; \pi(\vec\lambda_{\rm ex})\;
\mathcal{L}(d \mid \vec\lambda_{\rm ex}, \vec\lambda_{\rm in})\,,
\label{eq:ls}
\end{equation}
where $\vec\lambda_{\rm ex} \equiv (\hat n, \psi, \chi, \varphi_0, t, y)$ denote the extrinsic parameters (defined explicitly below), and $\vec\lambda_{\rm in}$ the intrinsic parameters.
We assume that the noise of each detector is stationary, Gaussian, and independent between the detectors, with one-sided power spectral density $S_I(f)$.
Thus, the noise variance of the data $\tilde d_{I,\nu}$ in frequency bin $\nu$ (at $f_\nu \equiv \nu / T_{\rm seg}$) is $\sigma^2_{I,\nu} = T_{\rm seg}\, S_I(f_\nu) / 2$.

\subsection{Rewriting the likelihood in terms of the filtered data and filter power}
\label{e:rewrite-filtered}

For a single source, the integrand of~\eqref{eq:ls} is a Gaussian likelihood, which in the frequency domain is a sum over the frequency bins of the two detectors,
\begin{equation}
\ln \frac{\mathcal{L}(d \mid \vec\lambda_{\rm ex}, \vec\lambda_{\rm in})}
{\mathcal{L}_n(d)}
= \sum_{I=1,2} \sum_\nu
\frac{2\, \mathrm{Re}[ \tilde d^{*}_{I,\nu} \tilde h_{I,\nu} ]
- |\tilde h_{I,\nu}|^2}{\sigma^2_{I,\nu}}\,.
\label{eq:lnratio}
\end{equation}
For the signal we use the two-detector waveform
\begin{equation}
\begin{aligned}
\tilde h_{I,\nu} &\equiv y\, R_I(\hat n, \psi, \chi)\, a_\nu(\vec\lambda_{\rm in})\,e^{-i\varphi_0}\\
&\hspace{0.75in} \times
e^{i\Psi_0(f_\nu, \vec\lambda_{\rm in})}\,
e^{-2\pi i f_\nu ( t + \tau_I(\hat n) )}\,.
\end{aligned}
\label{eq:template}
\end{equation}
Here $t$ is the arrival time at the second detector and $\tau_I(\hat n) \equiv \hat n \cdot (\vec x_I - \vec x_2)/c$ is the delay at detector $I$ for detector positions $\vec x_I$, so that $\tau_2 = 0$.
(In what follows, we define $\Delta\vec x \equiv \vec x_1 - \vec x_2$ and $\tau \equiv \tau_1= \hat n\cdot\Delta\vec x/c$.)
$y \equiv D_0/D$ is the amplitude scale of a source at luminosity distance $D$, relative to a fiducial distance $D_0$.
The response of detector $I$ to the sky direction $\hat n$, polarization angle $\psi$, and cosine of the inclination $\chi \equiv \cos\iota$ is
\begin{equation}
R_I(\hat n, \psi, \chi) = A_+(\chi)\,
F_I^+(\hat n, \psi) + i\, A_\times(\chi)\, F_I^\times(\hat n, \psi)\,,
\label{eq:response}
\end{equation}
where $F_I^+(\hat n, \psi)$ and $F_I^\times(\hat n, \psi)$ are the antenna pattern functions of the two detectors, and $A_+ \equiv(1 + \chi^2)/2$ and $A_\times \equiv\chi$ are the inclination weights of the dominant quadrupole mode.
The waveform enters through its amplitude $a_\nu(\vec\lambda_{\rm in})$ at distance $D_0$ and its phase $\Psi_0(f_\nu, \vec\lambda_{\rm in})$. The waveform~\eqref{eq:template} therefore has a fixed elliptical polarization, which holds for a nonprecessing binary whose signal keeps only the dominant quadrupole mode. This assumption separates the detector response from the frequency dependence in~\eqref{eq:linstep} and allows the analytic marginalization over the coalescence phase in~\eqref{eq:bessel}. Most observed binaries satisfy it to a good approximation, because the quadrupole mode dominates their signal near merger also when the orbit precesses.

In the linear term of~\eqref{eq:lnratio} the detector response, the coalescence phase, and the amplitude scale do not depend on frequency.
Hence, they can be taken out of the sum over frequency bins,
\begin{equation}
\begin{aligned}
\sum_\nu
\frac{2\, \mathrm{Re}[ \tilde d^{*}_{I,\nu} \tilde h_{I,\nu} ]}{\sigma^2_{I,\nu}}
&= 2 y\, \mathrm{Re}\Big[ R_I(\hat n, \psi, \chi)\,
e^{-i\varphi_0} \sum_\nu\!
\frac{\tilde d^{*}_{I,\nu}}{\sigma^2_{I,\nu}}\\
&\quad\times a_\nu(\vec\lambda_{\rm in}) e^{i \Psi_0(f_\nu, \vec\lambda_{\rm in})}
e^{-2\pi i f_\nu (t + \tau_I)}\Big]\,,
\end{aligned}
\label{eq:linstep}
\end{equation}
leaving a discrete Fourier series in the time $t + \tau_I$, since $f_\nu = \nu / T_{\rm seg}$.
In the quadratic term, every phase has unit modulus, leaving only the amplitudes,
\begin{equation}
\sum_\nu \frac{|\tilde h_{I,\nu}|^2}{\sigma^2_{I,\nu}}
= y^2\, |R_I(\hat n, \psi, \chi)|^2
\sum_\nu \frac{a_\nu(\vec\lambda_{\rm in})^2}{\sigma^2_{I,\nu}}.
\label{eq:quadstep}
\end{equation}
We call these two remaining frequency sums the
{\it filtered data} and the
{\it filter power} of detector $I$, defined by
\begin{equation}
\begin{aligned}
\mathcal{Z}_I(t, \vec\lambda_{\rm in}) &\equiv \sum_\nu
\frac{\tilde d^{*}_{I,\nu}\, a_\nu(\vec\lambda_{\rm in})\,
e^{i \Psi_0(f_\nu, \vec\lambda_{\rm in})}}{\sigma^2_{I,\nu}}\,
e^{-2\pi i \nu t / T_{\rm seg}}\,,\\
u_I(\vec\lambda_{\rm in}) &\equiv \sum_\nu
\frac{a_\nu(\vec\lambda_{\rm in})^2}{\sigma^2_{I,\nu}}\,.
\end{aligned}
\label{eq:zu}
\end{equation}
Up to normalization and complex conjugation, $\mathcal{Z}_I$ is the matched-filter output of compact binary searches and $u_I$ is the squared norm of its template, so the new ingredients here are the higher-order combinations of $\mathcal{Z}_I$ and their marginalized weights. The two filter powers differ when the two detectors have different noise spectra $S_I(f)$. With them the log likelihood ratio~\eqref{eq:lnratio} takes the simple form
\begin{equation}
\boxed{
\ln \frac{\mathcal{L}(d \mid \vec\lambda_{\rm ex}, \vec\lambda_{\rm in})}
{\mathcal{L}_n(d)}
= 2 y\, \mathrm{Re}\big[ e^{-i\varphi_0}\, \mathcal{Z} \big] - y^2 A\,,
\label{eq:reduced}
}
\end{equation}
where
\begin{equation}
\begin{aligned}
\mathcal{Z}(t, \hat n, \psi, \chi, \vec\lambda_{\rm in})
&\equiv R_1\, \mathcal{Z}_1(t + \tau(\hat n), \vec\lambda_{\rm in})
+ R_2\, \mathcal{Z}_2(t, \vec\lambda_{\rm in})\,,\\
A(\hat n, \psi, \chi, \vec\lambda_{\rm in})
&\equiv |R_1|^2\, u_1(\vec\lambda_{\rm in})
+ |R_2|^2\, u_2(\vec\lambda_{\rm in})\,,
\end{aligned}
\label{eq:ZA}
\end{equation}
are the coherent combination of the filtered data and the filter power from the two detectors.

\subsection{Marginalizing over the extrinsic parameters}
\label{s:marg-external}

We now marginalize the likelihood ratio over the extrinsic parameters, with isotropic priors on the sky position, polarization and inclination, a uniform prior on the coalescence phase, and priors on the distance and the arrival time given below.
For the distance marginalization, we use the prior uniform in volume, $p(y) = 3\, y_{\min}^3\, y^{-4}$ on $y \geq y_{\min}$, although other priors work in the same way.
The distance can also be kept as a parameter, in which case the integral over $y$ in~\eqref{eq:marg} is left out and the likelihood ratio and the GORFs carry $y$ as a further argument (Appendix~\ref{app:unbiased-pop-parameters}).
The arrival time $t$ is uniform in a window of half width $\Delta t$ around the segment center $t_c$.
But since in every analysis of this paper the window is the whole segment, we have $t_c = \Delta t = T_{\rm seg}/2$, so the analysis uses no trigger time.

Writing every integral explicitly, the likelihood ratio at fixed intrinsic parameters $\vec\lambda_{\rm in}$ is
\begin{equation}
\begin{aligned}
\Lambda(d\mid\vec\lambda_{\rm in})\equiv{}&
\frac{1}{4\pi} \int d^2\hat n\,
\frac{1}{2\Delta t} \int_{t_c - \Delta t}^{t_c + \Delta t} dt\,
\frac{1}{\pi} \int_0^\pi d\psi\,
\frac{1}{2} \int_{-1}^1 d\chi\\
&\times\int_{y_{\min}}^\infty dy\, p(y)\,
\frac{1}{2\pi} \int_0^{2\pi} d\varphi_0\\
&\times\exp\!\big( 2 y\, \mathrm{Re}[ e^{-i\varphi_0} \mathcal{Z} ] - y^2 A \big)\,.
\label{eq:marg}
\end{aligned}
\end{equation}
The coalescence phase enters~\eqref{eq:marg} only through $\mathrm{Re}[ e^{-i\varphi_0} \mathcal{Z} ]$, and $\mathcal{Z}$ itself is independent of $\varphi_0$.
So, the uniform phase integral gives a Bessel function~\cite{DLMF},
\begin{equation}
\frac{1}{2\pi} \int_0^{2\pi} d\varphi_0\,
e^{2 y\, \mathrm{Re}[ e^{-i\varphi_0} \mathcal{Z} ]} = I_0( 2 y |\mathcal{Z}| )\,.
\label{eq:bessel}
\end{equation}
After performing this integral, the data enter only through $|\mathcal{Z}|$, and the argument of the Bessel function is the product of two SNRs,

\begin{equation}
2 y |\mathcal{Z}| = \rho_{\rm opt}(y)\, \rho_{\rm mf}\,,
\end{equation}
where
\begin{equation}
\begin{aligned}
&\rho^2_{\rm opt}(y) \equiv 2 \sum_{I,\nu} \frac{|\tilde h_{I,\nu}|^2}{\sigma^2_{I,\nu}} = 2 y^2 A\,, \\
&\rho_{\rm mf}^2 \equiv 2\, \Big| \sum_{I,\nu} \frac{\tilde d^{*}_{I,\nu} \tilde h_{I,\nu}}{\sigma^2_{I,\nu}} \Big|^2
\Big/ \sum_{I,\nu} \frac{|\tilde h_{I,\nu}|^2}{\sigma^2_{I,\nu}} = \frac{2\,|\mathcal{Z}|^2}{A}\,.
\end{aligned}
\label{eq:snr}
\end{equation}
Here, $\rho_{\rm opt}(y)$ is the {\it optimal} network SNR of the source, and $\rho_{\rm mf}$ is the {\it matched-filter} network SNR of the data.

We then expand the Bessel function in its power series
\begin{equation}
I_0(2 y |\mathcal{Z}|) \simeq \sum_{n=0}^{n_{\max}}
\frac{(y |\mathcal{Z}|)^{2n}}{(n!)^2}\,.
\label{eq:taylor}
\end{equation}
As we show in Appendix~\ref{s:single}, after we  average over the distance to the source, the terms in the above expansion increase up to $n \approx \rho_{\rm mf}^2/2$, but then fall off beyond this limit.
So, for convergence, we need $n_{\max}$ to exceed $\rho_{\rm mf}^2/2$ for the loudest merger events we want to describe.
(See Appendix~\ref{app:BF-convergence} for a numerical test of this condition.)

We start with the first order, $n = 1$.
Its contribution to~\eqref{eq:marg} is the average of $y^2 |\mathcal{Z}|^2 e^{-y^2 A}$ over the extrinsic parameters.
Integrating over the distance gives the first distance moment, $M_1(A)$, where in general
\begin{equation}
M_n(A) \equiv \int_{y_{\min}}^\infty dy\, p(y)\, y^{2n}\, e^{-y^2 A}\,.
\label{eq:Mdist}
\end{equation}
By~\eqref{eq:ZA}, the absolute square of $\mathcal{Z}$ has four terms,
\begin{equation}
|\mathcal{Z}|^2 = |R_1|^2 |\mathcal{Z}_1|^2 + |R_2|^2 |\mathcal{Z}_2|^2 + R_1 R_2^* \mathcal{Z}_1 \mathcal{Z}_2^* + R_1^* R_2 \mathcal{Z}_1^* \mathcal{Z}_2\,.
\label{eq:binom1}
\end{equation}
Taking the cross term $R_1 R_2^*\, \mathcal{Z}_1 \mathcal{Z}_2^*$ as an example, its contribution to~\eqref{eq:marg} is
\begin{equation}
\begin{aligned}
&\frac{1}{4\pi}\!\int\! d^2\hat n\,
\frac{1}{\pi}\!\int_0^\pi\! d\psi\,
\frac{1}{2}\!\int_{-1}^{1}\! d\chi\;
M_1\big(A(\hat n, \psi, \chi,\vec\lambda_{\rm in})\big)\, R_1 R_2^*\\
&\hspace{0.5in}\times \frac{1}{2\Delta t}\!
\int_{t_c-\Delta t}^{t_c+\Delta t}\! dt\;
\mathcal{Z}_1(t + \tau(\hat n), \vec\lambda_{\rm in})\,
\mathcal{Z}_2^*(t, \vec\lambda_{\rm in})\,.
\label{eq:crossterm}
\end{aligned}
\end{equation}

The sky direction $\hat n$ appears above in both lines, through the responses and through the delay $\tau(\hat n)$ of $\mathcal{Z}_1$.
To separate these dependencies, we insert the identity $1 = \int d\tau\, \delta( \tau - \hat n\cdot\Delta \vec x / c )$ under the orientation average, and exchange the order of the integrals.
This leads to
\begin{align}
\text{Eq.}~\eqref{eq:crossterm}
&= \int\! d\tau
\Big\langle \delta\big( \tau -
\tfrac{\hat n \cdot \Delta \vec x}{c} \big)\,
M_1(A)\, R_1 R_2^* \Big\rangle_{\hat n, \psi, \chi}
\notag\\
&\;\times \frac{1}{2\Delta t}\!
\int_{t_c-\Delta t}^{t_c+\Delta t}\! dt\;
\mathcal{Z}_1(t + \tau, \vec\lambda_{\rm in})\, \mathcal{Z}_2^*(t, \vec\lambda_{\rm in}\,)\,,
\label{eq:splitorient}
\end{align}
where the orientation average is

\begin{equation}
\langle Q\rangle_{\hat n, \psi, \chi} \equiv \frac{1}{4\pi}\int d^2\hat n\, \frac{1}{\pi}\int_0^\pi d\psi\, \frac{1}{2}\int_{-1}^{1} d\chi\,Q(\hat n, \psi,\chi)
\label{eq:orient}
\end{equation}
for arbitrary $Q\equiv Q(\hat n,\psi,\chi)$.
The first line of \eqref{eq:splitorient} now depends only on the geometry, the distance prior, and the filter powers.
So we can compute it once and tabulate it as the weight $W^{(1)}_{10}(\tau, u_1, u_2)$, while the second line of \eqref{eq:splitorient} depends only on the data and the delay $\tau$, and we call it $T^{(1)}_{10}(\tau, \vec\lambda_{\rm in})$.

We can generalize the 1st-order calculation to every order since each factor of $\mathcal{Z}$ is the sum of a term of the first detector and a term of the second.
So, the binomial theorem gives
\begin{align}
|\mathcal{Z}|^{2n} = \sum_{a=0}^{n}\sum_{b=0}^{n}
\binom{n}{a}\binom{n}{b}\,
& R_1^{a} R_1^{*b} R_2^{n-a} R_2^{*(n-b)}\notag\\
\times\; \big[\mathcal{Z}_1^{a} \mathcal{Z}_1^{*b}\big](t + \tau(\hat n))\,
& \big[\mathcal{Z}_2^{n-a} \mathcal{Z}_2^{*(n-b)}\big](t)\,,
\label{eq:binomn}
\end{align}
where $a$ and $b$ count the factors of $\mathcal{Z}$ and $\mathcal{Z}^*$ taken from the first detector.
Using this expansion, the likelihood ratio then becomes
\begin{equation}
\boxed{
\begin{aligned}
\Lambda(\vec\lambda_{\rm in}) = \sum_{n=0}^{\infty} \frac{1}{(n!)^2}
\sum_{a=0}^{n} \sum_{b=0}^{n} & \binom{n}{a} \binom{n}{b} \\
\times \int d\tau\, & W^{(n)}_{ab}(\tau, u_1, u_2)\,
T^{(n)}_{ab}(\tau, \vec\lambda_{\rm in})\,,
\end{aligned}
}
\label{eq:master}
\end{equation}
which is the main result of this paper.
We denote its sum up to order $n_{\max}$ by $\Lambda_{n_{\max}}$. We compute the data factors $T^{(n)}_{ab}$ and the geometrical factors  $W^{(n)}_{ab}$ in the following subsections.
The series does not need to be evaluated term by term because at a given delay every order depends on the data only through the magnitudes $|\mathcal{Z}_1|$ and $|\mathcal{Z}_2|$ and their relative phase.
We therefore sum the orders once into a table of these three quantities (Appendix~\ref{app:construction}) so that analyzing a segment costs only the matched filters of its templates and one table lookup at every time and delay for any $n_{\max}$.

In Appendix~\ref{app:formalism}, we derive the same result from the characteristic function of the strain of Ref.~\cite{companion}, which makes explicit why the likelihood compresses into the statistics $T^{(n)}_{ab}$.
In Appendix~\ref{app:norm}, we show that adding a data-independent constant restores the unit noise mean of the truncated likelihood ratio, which we need only when $n_{\max}$ does not exceed $\rho_{\rm mf}^2/2$ for the loudest segment to which the expansion is applied.
This constant does not reproduce the omitted terms of a single segment, which are small only for $\rho_{\rm mf}^2/2 < n_{\max}$.

\subsection{Data factors}
\label{s:statistics}

After the marginalization over the extrinsic parameters, the likelihood ratio~\eqref{eq:master} depends on the data only via the combinations
\begin{equation}
\begin{aligned}
&T^{(n)}_{ab}(\tau, \vec\lambda_{\rm in}) = \big\langle \mathcal{Z}_1^{a} \mathcal{Z}_1^{*b}(t + \tau)\, \mathcal{Z}_2^{n-a} \mathcal{Z}_2^{*(n-b)}(t) \big\rangle_t \\
&\hspace{0.25in}\equiv \frac{1}{2\Delta t}
\int_{t_c - \Delta t}^{t_c + \Delta t} dt\,
\mathcal{Z}_1^{a} \mathcal{Z}_1^{*b}(t + \tau)\, \mathcal{Z}_2^{n-a} \mathcal{Z}_2^{*(n-b)}(t)\,.
\end{aligned}
\label{eq:T}
\end{equation}
Each such data factor is a delayed cross moment of the filtered data of the two detectors, averaged over the arrival window.
These data combinations compress the many frequency bins of each detector's data of a given segment into a set of functions of the delay without any loss for the marginalized likelihood.
They therefore play the role of {\it sufficient statistics} for the analysis.

\begin{figure*}[!tbp]
\centering
\includegraphics[width=\textwidth]{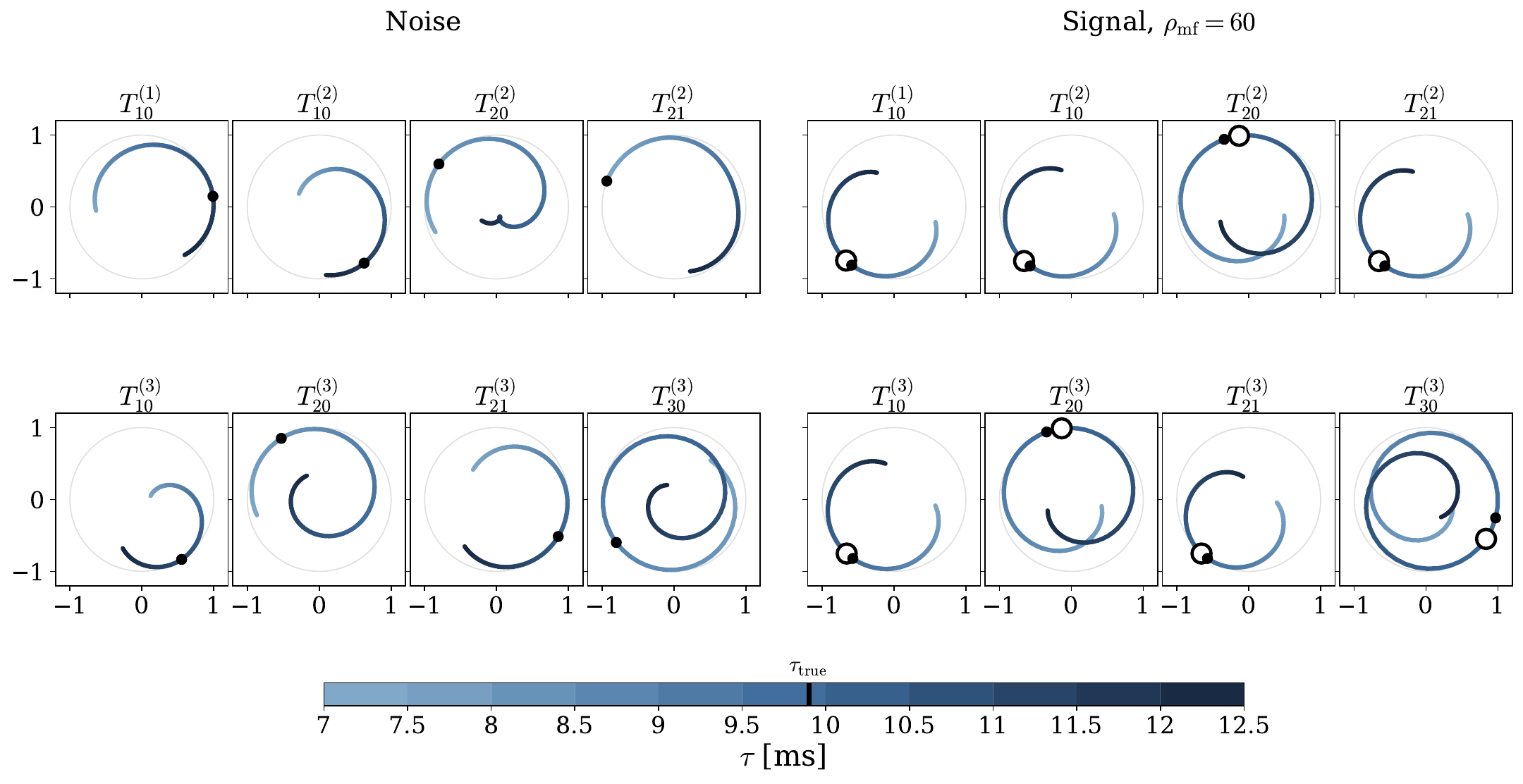}
\caption{The data factors of Eq.~\eqref{eq:master}.
Plotted are the normalized $T^{(n)}_{ab}(\tau, \vec\lambda_{\rm in})$ of \eqref{eq:T} for various orders $n$ and indices $a$, $b$ for a noise-only segment (left half) and for a segment containing an injected merger (right half), with each curve scaled to its own maximum.
The right-half curves are evaluated at the true intrinsic parameters, $\vec\lambda_{\rm in} = \vec\lambda_{{\rm in},{\rm true}}$.
The black dots mark the delay of each curve's maximum and the open circles mark the true delay, around which the dots scatter by the timing accuracy of the event.}
\label{fig:tdata}
\end{figure*}

Figure~\ref{fig:tdata} shows these statistics as curves in the complex plane parametrized by the delay, for a noise segment (left half) and for a segment with a loud merger (right half).
For the noise-only segment, the different statistics reach their maxima at unrelated delays.
For the segment with  a merger, all of the statistics reach their maxima near the true delay of the merger, with the phase of $T^{(n)}_{ab}$ winding $(a - b)$ times as fast as that of the cross-correlation $T^{(1)}_{10}$.

At first order, the statistics are the filtered auto powers of the two detectors and their filtered cross-correlation,
\begin{equation}
\begin{aligned}
T^{(1)}_{11} &= \langle |\mathcal{Z}_1|^2 \rangle_t\,, \\
T^{(1)}_{00} &= \langle |\mathcal{Z}_2|^2 \rangle_t\,, \\
T^{(1)}_{10}(\tau) &= \langle \mathcal{Z}_1(t + \tau)\, \mathcal{Z}_2^*(t) \rangle_t\,,
\end{aligned}
\label{eq:T1stats}
\end{equation}
where the first two do not depend on the delay when the window covers the whole segment.
These are filtered equivalents to the statistics of the standard stochastic searches---the excess power of each detector and the cross-correlation of the pair.
At second order, the series adds the fourth auto moments $\langle |\mathcal{Z}_I|^4 \rangle_t$ and the cross moments $T^{(2)}_{ab}(\tau)$ with the two simplest members
\begin{equation}
\begin{aligned}
T^{(2)}_{11}(\tau) &= \langle |\mathcal{Z}_1(t + \tau)|^2\, |\mathcal{Z}_2(t)|^2 \rangle_t\,, \\
T^{(2)}_{20}(\tau) &= \langle \mathcal{Z}_1^2(t + \tau)\, \mathcal{Z}_2^{*2}(t) \rangle_t\,.
\end{aligned}
\label{eq:T2stats}
\end{equation}
The first statistic is an intensity cross-correlation, while the second statistic correlates the two detectors at twice the phase of the chirp.

To see which information enters at each order, it is useful to write the time averages in the frequency domain.
We write the filtered data~\eqref{eq:zu} as $\mathcal{Z}_I(t, \vec\lambda_{\rm in}) = \sum_\nu c_{I,\nu}\, e^{-2\pi i \nu t / T_{\rm seg}}$, where $c_{I,\nu} \equiv \tilde d^{*}_{I,\nu}\, a_\nu(\vec\lambda_{\rm in})\, e^{i \Psi_0(f_\nu, \vec\lambda_{\rm in})} / \sigma^2_{I,\nu}$ is the contribution of bin $\nu$.
Since the filtered data are periodic with period $T_{\rm seg}$ and we average over the whole segment $2\Delta t = T_{\rm seg}$, the time average keeps only those terms whose frequencies add up to zero.
Thus, the first-order cross term becomes
\begin{equation}
T^{(1)}_{10}(\tau, \vec\lambda_{\rm in}) = \sum_\nu c_{1,\nu}\, c_{2,\nu}^*\, e^{-2\pi i \nu \tau / T_{\rm seg}}\,.
\label{eq:T1freq}
\end{equation}
Here, each bin is multiplied only by itself, so the phase $\Psi_0$ of the waveform cancels, and the first order statistic depends only on the cross power of each bin, similarly to the standard cross-correlation.
At second order, the same average gives, for example,
\begin{equation}
\begin{aligned}
T^{(2)}_{11}(\tau, \vec\lambda_{\rm in}) = {} & \sum_{\nu_1 \nu_2 \nu_3 \nu_4}
c_{1,\nu_1}\, c_{1,\nu_2}^*\, c_{2,\nu_3}\, c_{2,\nu_4}^* \\
& \times e^{-2\pi i (\nu_1 - \nu_2) \tau / T_{\rm seg}}\,
\delta_{\nu_1 - \nu_2 + \nu_3 - \nu_4,\, 0}\,,
\end{aligned}
\label{eq:T2freq}
\end{equation}
which multiplies four different bins, and now the phases $\Psi_0(f_{\nu_1}) - \Psi_0(f_{\nu_2}) + \Psi_0(f_{\nu_3}) - \Psi_0(f_{\nu_4})$ no longer cancel.
In the same way, the statistics of order $n$ correlate $2n$ frequency bins with each other. Statistics of a non-Gaussian background beyond the second moment of the data have been proposed before~\cite{Seto:2009ju,Martellini:2014xia,Buscicchio:2022raf}, and here every order follows from the marginalized likelihood, with its weight fixed by the GORFs.

\subsection{Geometrical factors}
\label{s:GORFs}

For the geometric factors, or GORFs, we average the detector responses over the orientation and the distance at a fixed delay,
\begin{equation}
\begin{aligned}
W^{(n)}_{ab}(\tau, u_1, u_2) = \Big\langle
& M_n(A)\, R_1^{a} R_1^{*b} R_2^{n-a} R_2^{*(n-b)} \\
& \times \delta\big( \tau - \tfrac{\hat n \cdot \Delta \vec x}{c} \big)
\Big\rangle_{\hat n, \psi, \chi},
\end{aligned}
\label{eq:W}
\end{equation}
with the orientation average of~\eqref{eq:orient} and the distance moment~\eqref{eq:Mdist}.
The Dirac delta function restricts the sky average to a ring of constant delay.
All waveforms with equal filter powers $u_1$ and $u_2$ therefore share the same GORFs.
The reflection $\chi \to -\chi$ maps every $R_I$ to $R_I^*$ and leaves $A$ and the prior unchanged.
The GORFs are therefore real and symmetric in $a$ and $b$.
When the distance is kept as a parameter (Sec.~\ref{s:marg-external}) the integral over $y$ is left out and $M_n(A)$ becomes $y^{2n} e^{-y^2 A}$ so that the GORFs carry $y$ as a further argument.
The population analysis of Sec.~\ref{sec:population} uses this form to keep the redshift of every merger as a parameter.

\begin{figure*}[!tbp]
\centering
\includegraphics[width=\textwidth]{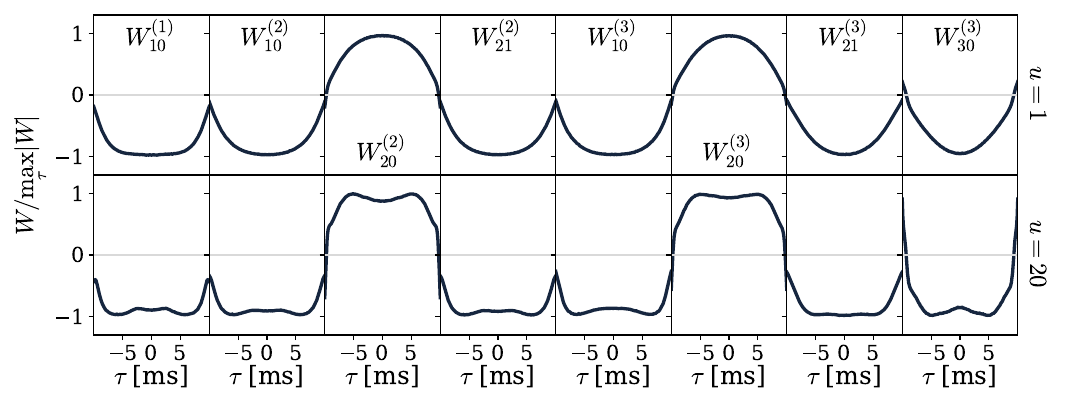}
\caption{The geometric factors of Eq.~\eqref{eq:master}.
Plotted are the normalized $W^{(n)}_{ab}(\tau, u_1, u_2)$ of \eqref{eq:W} for two values of the filter power (rows) under the distance prior that follows from the Madau--Dickinson rate history~\eqref{eq:rz} at the injected values of Appendix~\ref{app:pop-priors}.}
\label{fig:tables}
\end{figure*}

Figure~\ref{fig:tables} shows the GORFs for the LIGO Hanford and LIGO Livingston detectors for a small and large filter power $u\equiv u_1=u_2$.
For a small filter power, every GORF is a smooth function of the delay, while for a large filter power, they develop structure along the delay.
The orientation enters the distance moment~\eqref{eq:Mdist} only through the factor $e^{-y^2 A}$, which tends to $1$ on the whole ring as the filter powers go to zero.
In this limit, the GORFs reduce to averages of the responses over the ring.
However, for large filter power, $e^{-y^2 A}$ changes along the ring due to the response $A$, and the GORFs can then resolve the geometry of the source.
We note that a sky-averaged ORF inside a mixture likelihood biases the inferred duty cycle and amplitude of the background~\cite{Liu:2026xhc}, and the GORFs avoid this average by keeping the full dependence on the orientation.

It is also instructive to study the properties of the GORFs in the frequency domain, by taking their Fourier transform with respect to the delay $\tau$.
Since the delay between the detectors never exceeds the light travel time, every GORF vanishes outside $|\tau| \leq |\Delta\vec x|/c \approx 10$~ms.
Its Fourier transform is therefore a smooth function of frequency and oscillates with a period of about $c/|\Delta\vec x| \approx 100$~Hz.
The GORFs also jump to zero at the edges of the delay range, so their Fourier transforms fall off as $1/f$ at high frequency like the standard ORF.

\begin{figure*}[!tpb]
\includegraphics[width=\textwidth]{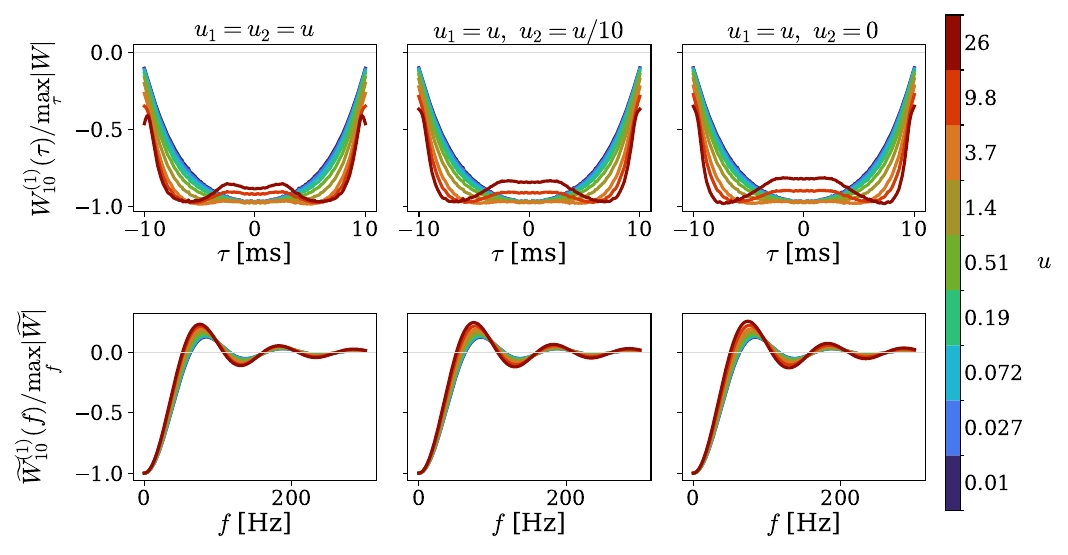}
\caption{The first order GORF in both the time and frequency domains.
Top panels: $W^{(1)}_{10}(\tau)$ for a set of filter powers (specified by the colorbar).
Bottom panels: The Fourier transform $\widetilde{W}^{(1)}_{10}(f)$ for the same set of filter powers.
The three columns correspond to a growing imbalance between the two filter powers, $u_1 = u_2 = u$, then $u_2 = u / 10$, then $u_2 = 0$.}
\label{fig:oru}
\end{figure*}

In the limit of zero filter power, $M_1(A)$ becomes the constant $M_1(0)$, which is finite for the prior uniform in volume, and the average of $R_1 R_2^*$ over $\psi$ and $\chi$ becomes proportional to $F_1^+ F_2^+ + F_1^\times F_2^\times$
evaluated at $\psi=0$~\cite{Allen1999,Romano2017}.
The transform of $W^{(1)}_{10}$ then reproduces the standard ORF $\Gamma^{(2)}_{HL}(f)$~\cite{Allen1999}, with its first zero near 64~Hz for the Hanford and Livingston pair,
\begin{equation}
\Gamma^{(2)}_{HL}(f) \propto \int d\tau\, e^{-2\pi i f \tau}\,
W^{(1)}_{10}(\tau, u_1, u_2)\big|_{u_1, u_2 \to 0}\,.
\label{eq:orf2}
\end{equation}
Figure~\ref{fig:oru} shows how this function changes as the filter power grows. In the columns with $u_2 < u_1$ the second detector is less sensitive, and the limit $u_2 \to 0$, reached for $\sigma^2_{2,\nu} \to \infty$, gives $A \to |R_1|^2 u_1$ in~\eqref{eq:ZA}.
The filter powers enter~\eqref{eq:W} only through $M_n(A)$, while the support $|\tau| \le |\Delta\vec x|/c$ of the Dirac delta is fixed by the baseline, so the Fourier transform oscillates with period $c/|\Delta\vec x|$ for every $(u_1, u_2)$.
At large power, its peaks change height and its zeros move.
So even the first order of the series $(n=1)$ depends on the SNR of the sources.

In the same way, the Fourier transforms of the second order GORFs give rise to the functions $\Gamma^{(4)}_{IJKL}(f_a, f_b, f_c, f_d)$, where the indices $IJKL$ label the detectors of the four factors $\mathcal{Z}_I \mathcal{Z}_J^* \mathcal{Z}_K \mathcal{Z}_L^*$.
The detector responses do not depend on frequency, so the frequencies enter only through the phase of the delay, for example,
\begin{equation}
\begin{aligned}
\Gamma^{(4)}_{HLHL}&(f_a, f_b, f_c, f_d) \\
&\propto \int d\tau\, e^{-\pi i (f_a + f_b + f_c + f_d) \tau}\, W^{(2)}_{20}(\tau, 0, 0)\,,
\end{aligned}
\label{eq:orf4}
\end{equation}
where we used $f_a - f_b + f_c - f_d = 0$ from the time average, and took the weak filter-power limit.%
\footnote{In this limit we replace the distance moment $M_n(A)$ by a constant, since $M_2(A)$ diverges as $A \to 0$ for a prior that is uniform in volume down to zero distance.}

\begin{figure*}[!tbp]
\centering
\includegraphics[width=\textwidth]{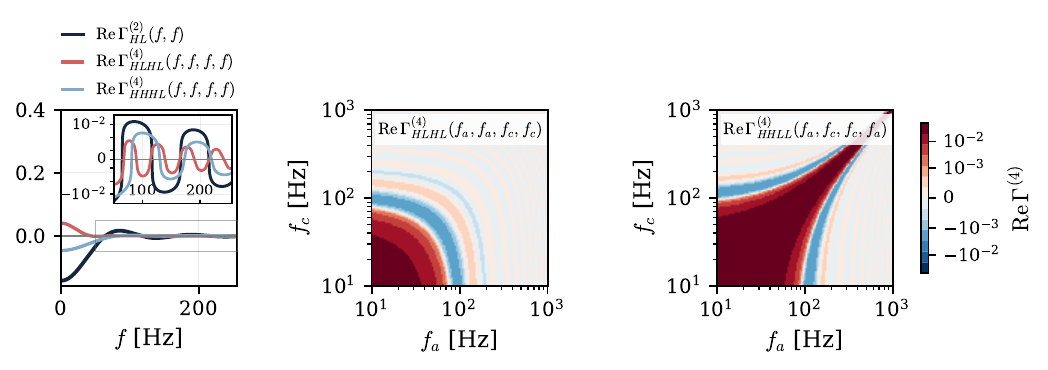}
\caption{GORFs of the LIGO Hanford and LIGO Livingston pair in the weak signal limit.
Left panel: the standard ORF $\Gamma^{(2)}_{HL}(f, f)$ and two equal frequency slices of $\Gamma^{(4)}$, with an inset above 50~Hz on a symmetric logarithmic axis.
Middle panel: $\Gamma^{(4)}_{HLHL}(f_a, f_a, f_c, f_c)$, which is a function of the sum $f_a + f_c$.
Right panel: $\Gamma^{(4)}_{HHLL}(f_a, f_c, f_c, f_a)$, which is a function of the difference $f_a - f_c$.
Both slices satisfy the frequency constraint of~\eqref{eq:T2freq}.}
\label{fig:orf4}
\end{figure*}

Figure~\ref{fig:orf4} shows the higher-order frequency-domain GORFs $\Gamma^{(4)}$ for different pairings of the frequency arguments, for the LIGO Hanford and LIGO Livingston pair.
On the equal frequency slices of the left panel, the delay phases of a factor $\mathcal{Z}_H$ and a factor $\mathcal{Z}_H^*$ cancel.
So $\Gamma^{(4)}_{HLHL}(f,f,f,f)$, with two factors of $\mathcal{Z}_H$, oscillates at twice the frequency of the ORF $\Gamma^{(2)}_{HL}(f,f)$, and $\Gamma^{(4)}_{HHHL}(f,f,f,f)$, with $\mathcal{Z}_H \mathcal{Z}_H^* \mathcal{Z}_H$, at the same frequency.
The middle panel shows that $\Gamma^{(4)}_{HLHL}(f_a,f_a,f_c,f_c)$ depends on the two frequencies $f_a$ and $f_c$ only via their sum.
So it correlates pairs of frequency bins whose phases add up.
The right panel shows that the intensity term $\Gamma^{(4)}_{HHLL}(f_a,f_c,f_c,f_a)$ depends on $f_a$ and $f_c$ only via their difference $f_a - f_c$.
So, this function is largest along the diagonal $f_a=f_c$, where the two frequency bins are close to one another.

In Appendix~\ref{app:construction}, we illustrate the convergence of the first-order GORF with the resolution of the sky grid.

\section{Validating the search on simulated data}
\label{sec:population}

\subsection{Simulation details}
\label{s:simulation}

To test the search on a realistic background, we simulate a year of data in the two LIGO detectors and divide it into segments of duration of $T_{\rm seg} = 8$~s.
For the noise we use a power spectral density estimated from public O4 data\footnote{The power spectral density is a median Welch average over blocks of 8~s of the public LIGO Livingston strain around GW250207\_115645~\cite{LIGOScientific:2026qqv} from the Gravitational Wave Open Science Center~\cite{GWOSC,LIGOScientific:2025snk,LIGOScientific:2026jgl}.
We use the same power spectral density for both detectors.} in the band from 20 to 512~Hz. We draw every segment directly in the frequency domain, with stationary Gaussian noise of this spectrum in each detector.
The binary black holes merge at a rate density of $R_0 = 19.1~{\rm Gpc^{-3}\,yr^{-1}}$ at $z = 0$~\cite{GWTC5}.
We assume a simple setup in which every merger is an equal-mass nonspinning binary modeled with the IMRPhenomXAS waveform~\cite{XAS}, and the analysis uses the same equal-mass nonspinning waveforms.
Its primary mass follows the default mass model of GWTC-5~\cite{GWTC5}, a broken power law with two Gaussian peaks, carried to the detector frame over a Madau--Dickinson redshift distribution~\cite{MadauDickinson2014} out to $z = 10$.
We choose the segment duration of 8~s because an equal-mass binary with a detector-frame total mass of $20\,M_\odot$ stays above 20~Hz for about 6~s and therefore fits inside one segment, and we do not simulate lighter binaries.
The sky position, orientation, phase and arrival time of the mergers are drawn from uniform distributions.

We simulate every segment independently with at most one merger inside it, so no chirp crosses a segment boundary and a segment contains a merger with probability $\bar N \simeq 0.019$.
We evaluate the likelihood ratio~\eqref{eq:master} for $n_{\max}$ up to 120 with the GORFs at the fixed distance $y = D_0/D(z)$ so that the mass and the redshift of the merger stay parameters.
In these GORFs, the distance moment $M_n(A)$ of~\eqref{eq:Mdist} is replaced by $y^{2n} e^{-y^2 A}$ as in Sec.~\ref{s:GORFs} (see also Appendix~\ref{app:unbiased-pop-parameters}). We infer seven population parameters: the rate density $R_0$, the three parameters $\alpha_1$, $\alpha_2$, and $\mu_2$ of the GWTC-5 mass model, and the three parameters $\gamma$, $\kappa$, and $z_p$ of the Madau--Dickinson rate history.
We denote the six parameters other than $R_0$ by $\vec\theta = (\alpha_1, \alpha_2, \mu_2, \gamma, \kappa, z_p)$.
The population model and the priors are described in Appendix~\ref{app:pop-priors}.

\subsection{Catalog-informed and uninformed searches}
\label{s:searches}

The data of current detectors already contain individually detectable mergers and a search for the background can therefore be done either without or with the information of the catalog of these mergers.
We then separate the segments with detectable mergers from the rest in the analysis.
We use $\rho_{\rm th}$ as the detection threshold on the largest matched-filter SNR $\rho_{\rm mf}^{\max}$ of a segment~\eqref{eq:snr}.
In our simulation, 467 segments are above $\rho_{\rm th} = 8$ and $3.9 \times 10^6$ segments are below it, and we denote the detected segments by $\mathcal{D}$ and the unresolved segments by $\mathcal{U}$.
We analyze the detected segments separately with a full parameter estimation of each segment that uses the exact likelihood.
We apply the expansion~\eqref{eq:master} truncated at $n_{\max}$ only to the unresolved segments.
Since the expansion converges for $n_{\max} > \rho_{\rm mf}^2/2$ (Sec.~\ref{sec:method}) and no unresolved segment exceeds $\rho_{\rm mf}^{\max} = \rho_{\rm th}$, the unresolved segments need a minimum order $n_{\min} = \rho_{\rm th}^2/2 = 32$.
In our analysis we use various $n_{\max}$ below and above $n_{\min}$ to test it.

The exact likelihood of the detected segments and the truncated expansion of the unresolved segments both enter the inference of the population through a hierarchical likelihood~\cite{Mandel:2018mve,Thrane:2018qnx}.
For a population with rate density $R_0$ and parameters $\vec\theta$, a segment contains a merger with probability $\bar N(R_0, \vec\theta)$, and the detector-frame total mass $M^{\rm det}$ and the redshift $z$ of the merger follow the distribution $\pi(M^{\rm det}, z \mid \vec\theta)$, both given in Appendix~\ref{app:pop-priors}.
Here $M^{\rm det} = 2 m_1^{\rm src} (1 + z)$ and $m_1^{\rm src}$ is the source-frame primary mass.
We average the likelihood ratio of every segment $i$ over this distribution,
\begin{equation}
\bar\Lambda_i(\vec\theta) = \int_0^{10}\! dz \int_{20\,M_\odot}^{1000\,M_\odot}\!\! dM^{\rm det}\, \pi(M^{\rm det}, z \,|\, \vec\theta) \Lambda_i(M^{\rm det}, z),
\label{eq:lambar}
\end{equation}
where
\begin{equation}
\Lambda_i = \begin{cases}
\big\langle e^{2 y\, \mathrm{Re}[ e^{-i\varphi_0} \mathcal{Z} ] - y^2 A} \big\rangle_{\hat n, t, \psi, \chi, \varphi_0}\,, & i \in \mathcal{D}\,, \\[2pt]
\Lambda_{n_{\max}}~\text{of~\eqref{eq:master}}\,, & i \in \mathcal{U}\,.
\end{cases}
\label{eq:lamcases}
\end{equation}
For the detected segments, we calculate the exact likelihood ratio~\eqref{eq:marg} at the distance $y = D_0/D(z)$, and for the unresolved segments, the expansion truncated at $n_{\max}$.
Both are evaluated on one grid of masses and redshifts common to all the segments (Appendix~\ref{app:construction}), so the integral~\eqref{eq:lambar} is a sum over this grid and carries no sampling uncertainty.
Searches that rely on stochastic sampling estimate the same average from a finite number of posterior samples of every segment~\cite{Smith2018,Thrane:2018qnx}.
The uncertainty of this Monte Carlo estimate grows with the number of segments~\cite{Talbot:2023pex}, and Ref.~\cite{Bers:2025tei} found that it biases the duty cycle and the redshift distribution, which they avoid by likelihood reweighting.

The likelihood of the catalog and the likelihood of the unresolved segments, both relative to the noise-only model, are
\begin{align}
\mathcal{L}_{\mathcal{D}}(R_0, \vec\theta) &= \Big[ 1 - \bar N(R_0, \vec\theta)\, f_{\rm det}(\vec\theta) \Big]^{N_{\mathcal{U}}} \nonumber \\
&\quad \times \prod_{i \in \mathcal{D}} \Big[ 1 + \bar N(R_0, \vec\theta)\, \big( \bar\Lambda_i(\vec\theta) - 1 \big) \Big]\,,
\label{eq:LD} \\
\mathcal{L}_{\mathcal{U}}(R_0, \vec\theta) &= \Big[ 1 - \bar N(R_0, \vec\theta)\, f_{\rm det}(\vec\theta) \Big]^{-N_{\mathcal{U}}} \nonumber \\
&\quad \times \prod_{i \in \mathcal{U}} \Big[ 1 + \bar N(R_0, \vec\theta)\, \big( \bar\Lambda_i(\vec\theta) - 1 \big) \Big]\,,
\label{eq:LU}
\end{align}
where $N_{\mathcal{U}}$ is the number of unresolved segments and
\begin{align}
f_{\rm det}(\vec\theta) &= \int_0^{10}\! dz \int_{20\,M_\odot}^{1000\,M_\odot}\!\! dM^{\rm det}\; \pi(M^{\rm det}, z \mid \vec\theta) \nonumber \\
&\quad \times P\big(\rho_{\rm mf}^{\max} > \rho_{\rm th} \mid M^{\rm det}, z\big)
\label{eq:alpha}
\end{align}
is the fraction of the mergers of the population that are detected, with $P$ the probability that a merger of mass $M^{\rm det}$ at redshift $z$ is detected (Appendix~\ref{app:pop-priors}).
The first factor of $\mathcal{L}_{\mathcal{D}}$ is the probability that the unresolved segments contain no detectable merger, and $\mathcal{L}_{\mathcal{U}}$ divides by it to account for the selection of these segments below the threshold~\cite{Mandel:2018mve}.
The two factors cancel in the product $\mathcal{L}_{\mathcal{D}}\, \mathcal{L}_{\mathcal{U}}$, which is the mixture likelihood~\eqref{eq:mixture2} of all the segments.

We search for the background in the unresolved segments with two priors on the population.
The \emph{uninformed search} does not use the catalog and its prior is $\pi_0$ of~\eqref{eq:pi0}, which is uniform in $\ln R_0$ and in the six other parameters.
In the uninformed search, this selection factor does not cancel because only the unresolved segments enter its likelihood.
Its Bayes factor against the noise-only model and its posterior are
\begin{align}
&\mathcal{B}_{\rm U} = \int d\ln R_0\, d\vec\theta\; \pi_0(R_0, \vec\theta)\, \mathcal{L}_{\mathcal{U}}(R_0, \vec\theta)\,,
\label{eq:BU} \\
&p_{\rm U}(R_0, \vec\theta) = \frac{\pi_0(R_0, \vec\theta)\, \mathcal{L}_{\mathcal{U}}(R_0, \vec\theta)}{\mathcal{B}_{\rm U}}\,.
\label{eq:pU}
\end{align}
The \emph{catalog-informed search} uses the posterior of the catalog $\pi_{\mathcal{D}}(R_0, \vec\theta) \propto \pi_0(R_0, \vec\theta)\, \mathcal{L}_{\mathcal{D}}(R_0, \vec\theta)$, normalized over the seven parameters, as its prior,
\begin{align}
&\mathcal{B}_{\rm I} = \int d\ln R_0\, d\vec\theta\; \pi_{\mathcal{D}}(R_0, \vec\theta)\, \mathcal{L}_{\mathcal{U}}(R_0, \vec\theta)\,,
\label{eq:BI} \\
&p_{\rm I}(R_0, \vec\theta) = \frac{\pi_{\mathcal{D}}(R_0, \vec\theta)\, \mathcal{L}_{\mathcal{U}}(R_0, \vec\theta)}{\mathcal{B}_{\rm I}}\,.
\label{eq:pI}
\end{align}
Since $p_{\rm I} \propto \pi_0\, \mathcal{L}_{\mathcal{D}}\, \mathcal{L}_{\mathcal{U}}$, the posterior of the catalog-informed search equals the posterior of all the segments under $\pi_0$.
In the rest of this section, every prior is either $\pi_0$ or $\pi_{\mathcal{D}}$, and we name the search that uses it.

We confirm in Appendix~\ref{app:other-cases} that for $\rho_{\rm th} = 8$ the Bayes factors and the posteriors of both searches are the same for the three orders above $n_{\min}$ ($n_{\max} = 40$, 80, and 120) and differ for the order below it ($n_{\max} = 20$).
We also repeat the two searches assuming that only the segments with $\rho_{\rm mf}^{\max} > 12$ are detectable, and we also analyze all the segments with the truncated expansion without separating the detectable ones.

\subsection{Posterior distributions for the population parameters}
\label{s:popparms-main}

\begin{figure*}[!htbp]
\centering
\includegraphics[width=0.8\textwidth]{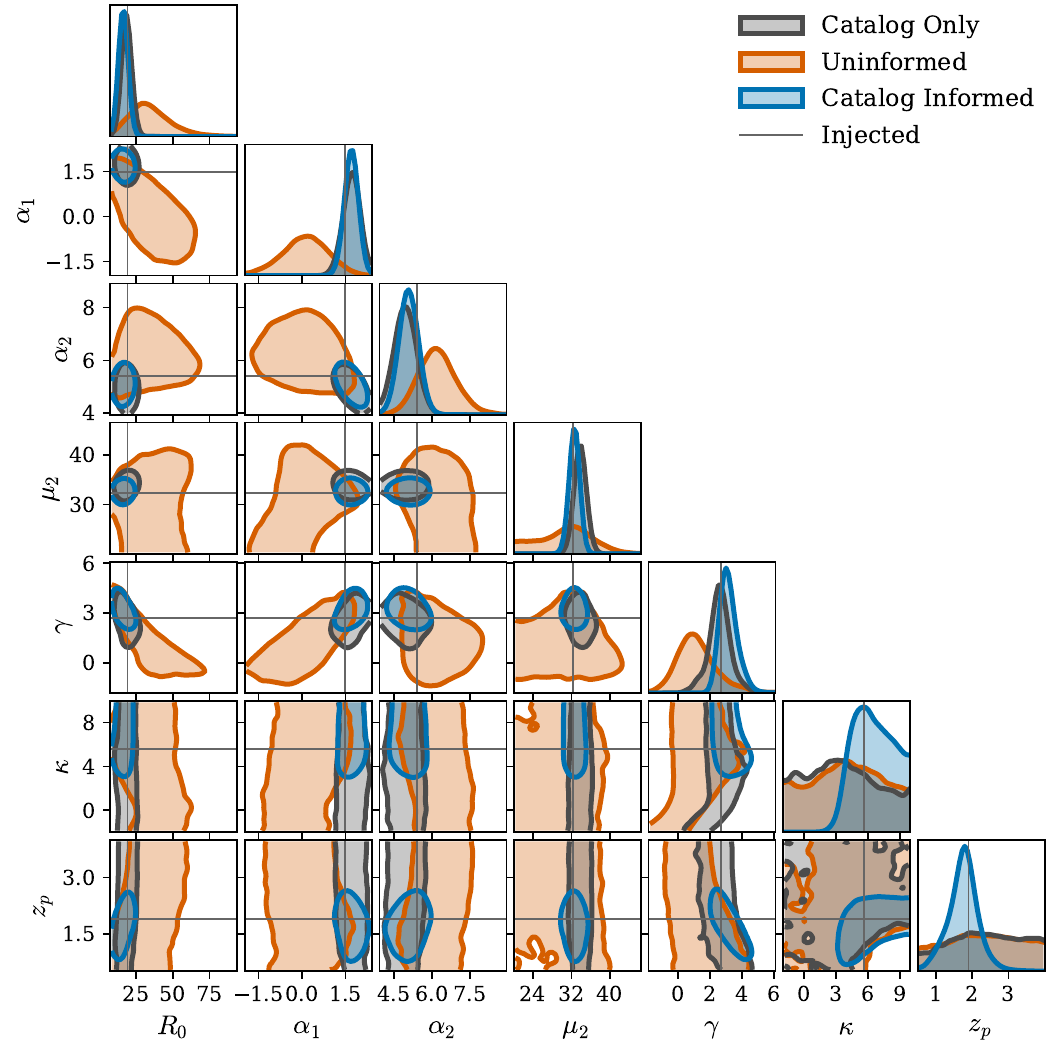}
\caption{Posteriors of the rate density $R_0$ in ${\rm Gpc^{-3}\,yr^{-1}}$ and the six parameters of the mass model and the rate history, with $\mu_2$ in $M_\odot$, for $\rho_{\rm th} = 8$ and $n_{\max} = 120$.
The grey regions show the posteriors $\pi_{\mathcal{D}}$ for the detected segments alone.
The colored regions show the posteriors for the uninformed search (orange), in which the segments above the threshold are left out and the prior is $\pi_0$, and for the catalog-informed search (blue), in which the prior is $\pi_{\mathcal{D}}$.
The contours enclose the 90 percent credible regions, and the grey lines mark the injected values.}
\label{fig:pop}
\end{figure*}

We compare the population inferred from the catalog only, from the unresolved segments only, and from the two together.
Figure~\ref{fig:pop} shows the posteriors of the seven population parameters.
The grey regions show the posterior $\pi_{\mathcal{D}}$ of the catalog that serves as the prior of the catalog-informed search, and the colored regions show the posteriors $p_{\rm U}$ of the uninformed search and $p_{\rm I}$ of the catalog-informed search.

Figure~\ref{fig:pop} shows that all three posteriors are consistent with the injected values except the low-mass slope $\alpha_1$ in the uninformed search that we discuss below.
For the local rate $R_0$ and the four parameters $\alpha_1$, $\alpha_2$, $\mu_2$, and $\gamma$ in the upper rows, the three posteriors differ mostly in their width.
The uninformed search gives the widest posteriors, so the unresolved segments alone constrain these parameters without any detected merger, but less tightly than the catalog.
The posteriors of the catalog and of the catalog-informed search nearly coincide, so the unresolved segments add little to the catalog for these parameters and the local rate stays at $R_0 = 16.2^{+4.3}_{-4.0}~{\rm Gpc^{-3}\,yr^{-1}}$ (median and 90 percent interval).
For the parameters $\kappa$ and $z_p$ of the peak of the rate history in the lower two rows, the conclusion is the opposite.
The catalog-only and the uninformed posteriors leave these two parameters at their prior, and only the two together constrain the peak to $z_p = 1.77^{+0.65}_{-0.74}$.
\begin{figure*}
\includegraphics[width=0.8\linewidth]{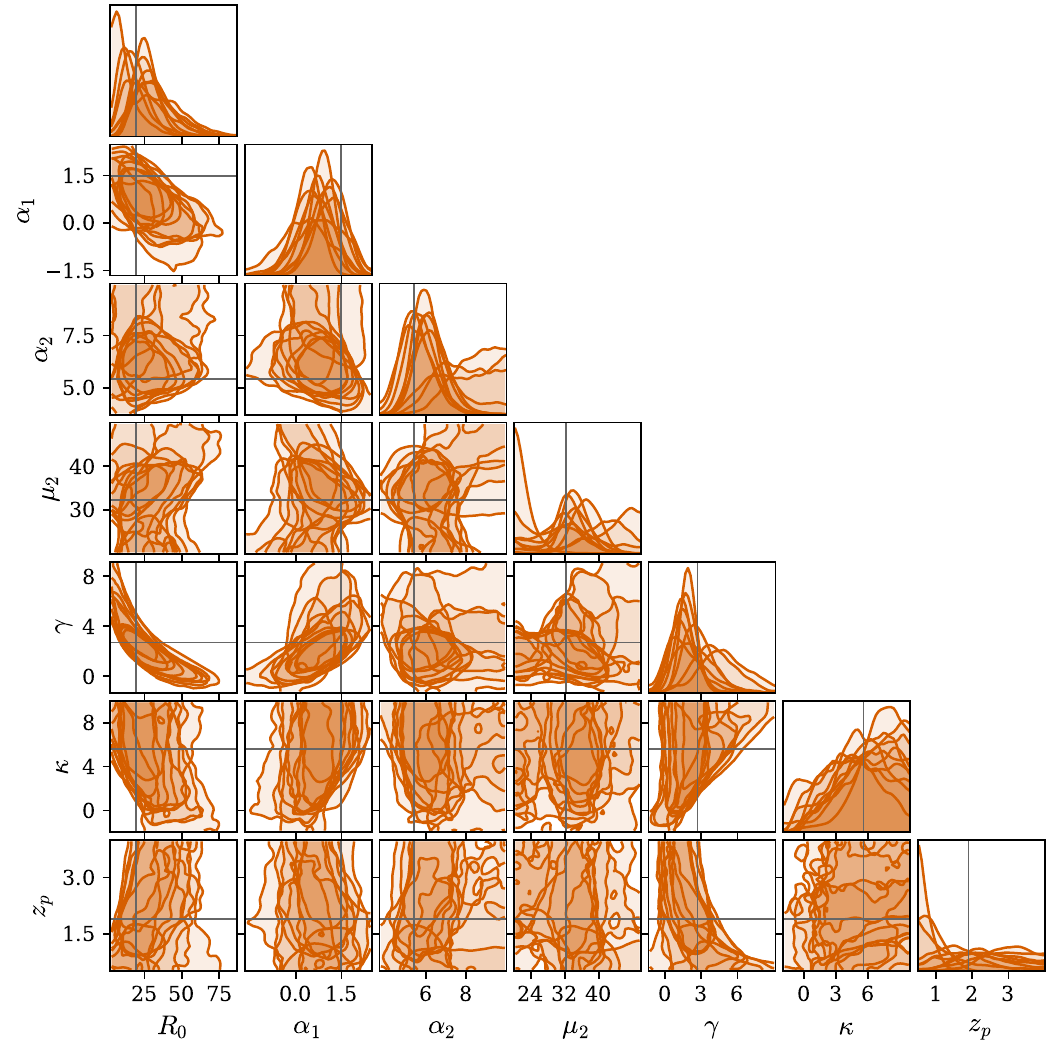}
\caption{Posteriors of the uninformed search for $\rho_{\rm th} = 8$ and $n_{\max} = 120$ in ten independent realizations of the simulated data, with 90 percent credible regions. The grey lines mark the injected values.}
\label{fig:pop_years}
\end{figure*}

In the uninformed search, the posteriors of $R_0$, $\gamma$, and $\alpha_1$ are degenerate along a direction in which a larger local rate is compensated by a slower rise of the rate with redshift and by fewer light binaries.
Its marginal posteriors are therefore shifted toward a larger $R_0$ and a smaller $\gamma$ and $\alpha_1$, although the injected values lie inside its two-dimensional contours.
To test whether this shift is a fluctuation of one realization we repeat the uninformed search for ten independent realizations of the simulated data (Fig.~\ref{fig:pop_years}) and find that the injected $R_0$, $\gamma$, and $\kappa$ lie inside the 90 percent intervals for all ten years with posterior medians on both sides of the injected values.
The posterior median of the low-mass slope $\alpha_1$ however lies below its injected value in all ten years and that of the high-mass slope $\alpha_2$ above it, so the tilt of the mass distribution in the uninformed search is systematic.
The unresolved segments alone also locate the second peak of the mass distribution because the posterior of $\mu_2$ peaks at its injected value, although none of these segments contains a detectable merger.

\begin{figure*}[!htbp]
\centering
\includegraphics[width=\textwidth]{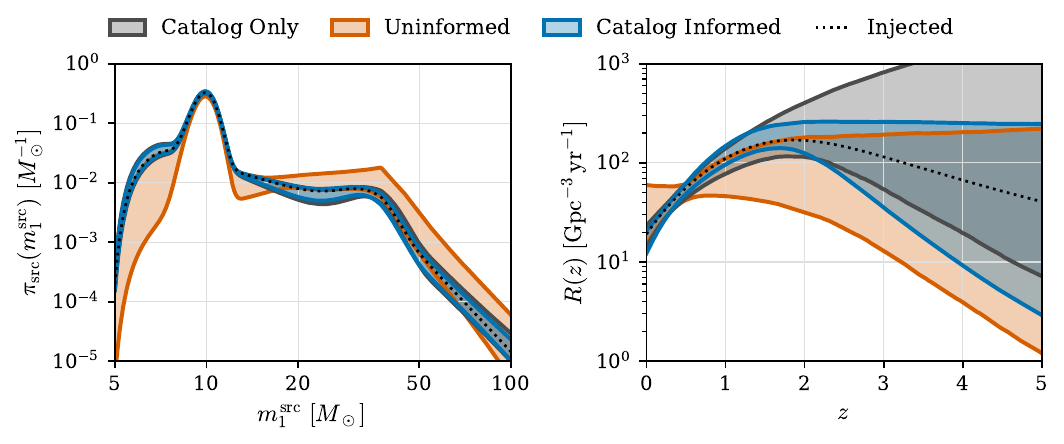}
\caption{The distribution of the source-frame primary mass $m_1^{\rm src}$ (left) and the merger rate density $R(z)$ of~\eqref{eq:rz} (right) over the posteriors of Fig.~\ref{fig:pop}, as bands between the 5th and the 95th percentile at every mass and at every redshift.
The dotted curves show the injected population.
Note that our population model fits only $\alpha_1$, $\alpha_2$ and $\mu_2$ of the mass distribution (Appendix~\ref{app:pop-priors}), so the peak at $10\,M_\odot$ keeps its injected shape in every band.
}
\label{fig:rz}
\end{figure*}

Figure~\ref{fig:rz} shows the same three posteriors as bands of the distribution $\pi_{\rm src}(m_1^{\rm src} \mid \vec\theta)$ of the source-frame primary mass (left) and of the merger rate density $R(z) = R_0\, \psi(z \mid \vec\theta)$ of~\eqref{eq:rz} (right).
In the left panel, all three bands follow the injected mass distribution and the band of the uninformed search is the widest.
This band lies above the injected distribution between the two peaks and below it at the lowest masses, because the uninformed search prefers a smaller $\alpha_1$ (Fig.~\ref{fig:pop}).
In the right panel, the catalog-only band constrains the rate below $z \simeq 1$ but leaves it unbounded above $z \simeq 2$.
The uninformed search constrains the rate best near $z \simeq 0.5$ and bounds it from above at all higher redshifts.
Its band widens toward $z = 0$ because a larger local rate with a slower rise fits the data equally well, and this is the degeneracy of the uninformed posterior along $R_0$ and $\gamma$ in Fig.~\ref{fig:pop}.
The catalog-informed band follows the catalog-only band at low redshift and the upper bound of the uninformed band at high redshift, so that the rate turns over near the injected peak.
Beyond the peak, this band still spans about two orders of magnitude, and the posterior of $\kappa$ reaches the upper edge of its prior.
The two data sets together therefore determine the existence and the location of the peak, but they are unable to constrain the falloff of the rate after the peak.
In Appendix~\ref{app:null}, we analyze several years of noise-only realizations. They are consistent with the null hypothesis, and the rate posterior peaks at the lower edge of its prior.

\subsection{Bayes factor distributions and time-to-detection}
\label{s:foreground-background}
\label{s:time-to-detection}

The time needed to detect the background is set by how often a segment with a large Bayes factor arrives.
We therefore look at the distribution of the Bayes factors of the unresolved segments in the uninformed and catalog-informed searches
\begin{align}
\mathcal{B}_i^{\rm U} &= \frac{\int d\ln R_0\, d\vec\theta\; \pi_0(R_0, \vec\theta)\, \bar\Lambda_i(\vec\theta)}{1 - \int d\ln R_0\, d\vec\theta\; \pi_0(R_0, \vec\theta)\, f_{\rm det}(\vec\theta)}\,,
\label{eq:segU} \\
\mathcal{B}_i^{\rm I} &= \frac{\int d\ln R_0\, d\vec\theta\; \pi_{\mathcal{D}}(R_0, \vec\theta)\, \bar\Lambda_i(\vec\theta)}{1 - \int d\ln R_0\, d\vec\theta\; \pi_{\mathcal{D}}(R_0, \vec\theta)\, f_{\rm det}(\vec\theta)}\,,
\label{eq:segI}
\end{align}
with $\bar\Lambda_i(\vec\theta)$ of~\eqref{eq:lambar} and the detected fraction $f_{\rm det}(\vec\theta)$ of~\eqref{eq:alpha}.
The numerator is the likelihood ratio of the segment averaged over the populations of the prior, and the denominator accounts for the detectable mergers of these populations, which are analyzed separately.
The two Bayes factors differ only in the prior, $\pi_0$ for the uninformed search and the posterior $\pi_{\mathcal{D}}$ of the catalog for the catalog-informed search.

\begin{figure}[!htbp]
\centering
\includegraphics[width=\columnwidth]{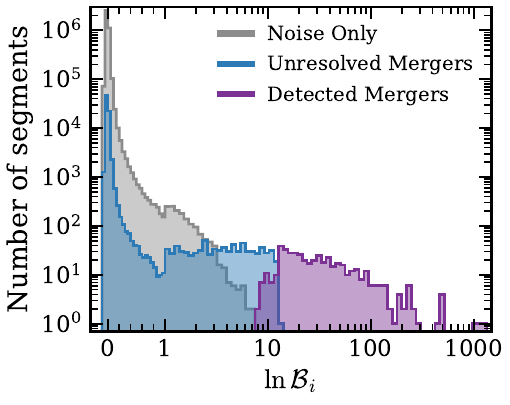}
\caption{Number of segments in bins of the log Bayes factor $\ln\mathcal{B}_i^{\rm I}$ of the catalog-informed search~\eqref{eq:segI} for $n_{\max} = 120$.
The unresolved segments are split into those with a merger and the noise-only segments, which is known only for the simulation.
For the detected segments, $\mathcal{B}_i$ is the exact likelihood ratio~\eqref{eq:marg} averaged over the same prior.
Since the threshold applies to $\rho_{\rm mf}^{\max}$ and not to $\mathcal{B}_i$, the unresolved and the detected mergers overlap.}
\label{fig:year_lnb}
\end{figure}

Figure~\ref{fig:year_lnb} shows the numbers of noise-only segments, unresolved mergers, and detected mergers at every value of $\ln\mathcal{B}_i$.
Most unresolved mergers follow the noise-only distribution and only $\mathcal{O}(100)$ of them exceed the largest value $\ln\mathcal{B}_i = 7.0$ that the noise-only segments reach.
The $\mathcal{O}(400)$ unresolved mergers with $\ln\mathcal{B}_i > 3$ are less than one percent of the total number of unresolved mergers for the full year, and contribute most of the log likelihood ratio of the unresolved segments (Appendix~\ref{app:other-cases}). (They arrive on average about once per day.)
The Bayes factors $\mathcal{B}_i^{\rm U}$ of the uninformed search~\eqref{eq:segU} have nearly the same distribution.
The distribution is also the same for the two orders above $n_{\min}$ ($n_{\max} = 40$ and 120), and Appendix~\ref{app:other-cases} shows the effect of the orders below $n_{\min}$ on the total Bayes factor and on the posteriors.

From Fig.~\ref{fig:year_lnb}, we also see that
the number of noise-only segments falls exponentially with $\ln\mathcal{B}_i$.
For large $|\mathcal{Z}|$, the integrand of the distance integral in~\eqref{eq:marg}, which the redshift integral of~\eqref{eq:lambar} replaces here, behaves as $e^{2y|\mathcal{Z}| - y^2 A}$.
It peaks at $y = |\mathcal{Z}|/A$, so $\ln \Lambda \simeq |\mathcal{Z}|^2/A$.
Under noise, $\mathcal{Z}$ is a complex Gaussian variable with $\mathbb{E}_0[|\mathcal{Z}|^2] = A$, so $|\mathcal{Z}|^2/A$ is exponentially distributed with unit mean.
The exponential tail also implies that the fraction of noise-only segments above $\ln\mathcal{B}_i$ falls roughly as $1/\mathcal{B}_i$.

\begin{figure*}[!htbp]
\centering
\includegraphics[width=0.8\textwidth]{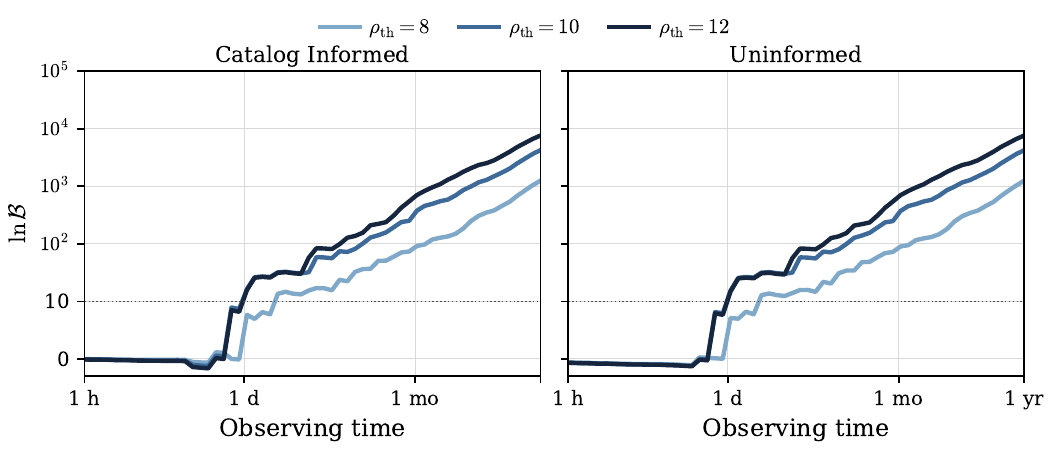}
\caption{Log Bayes factor of the catalog-informed search (left, $\ln\mathcal{B}_{\rm I}$) and the uninformed search (right, $\ln\mathcal{B}_{\rm U}$) against observing time for three values of $\rho_{\rm th}$ at $n_{\max} = 120$.
The dotted horizontal line marks $\ln\mathcal{B} = 10$, below which the vertical axis is linear and above which it is logarithmic.}
\label{fig:time_th}
\end{figure*}

Figure~\ref{fig:time_th} shows the log Bayes factors $\ln\mathcal{B}_{\rm I}$ and $\ln\mathcal{B}_{\rm U}$ of the two searches against observing time for three values of $\rho_{\rm th}$.
At every observing time, the detected segments are those that exceeded $\rho_{\rm th}$ up to that time, and the prior $\pi_{\mathcal{D}}$ of the catalog-informed search is the posterior of the catalog of that time.
For $\rho_{\rm th} = 8$, the log Bayes factors stay near zero during the first day and pass $\ln\mathcal{B} = 10$ after about two days.
They rise in steps as the first segments from the tail of the distribution of $\mathcal{B}_i$ arrive, and they grow almost in proportion to the observing time once many such segments have arrived.
A higher threshold moves the loudest segments from the catalog to the unresolved segments, so the background is detected earlier and its log Bayes factor after one year is larger.
At every threshold the two searches give nearly the same curve, so the catalog as prior does not make the detection faster, and the evidence comes from the unresolved segments themselves. With the catalog as prior, the posteriors of Fig.~\ref{fig:pop} become narrower and the detection time stays the same.
The curves for $n_{\max} = 40$ and 120 coincide (Fig.~\ref{fig:time5} of Appendix~\ref{app:other-cases}), so a larger $n_{\max}$ does not change the result above $n_{\min}$.
At low order, the likelihood ratio of any merger has a maximum value that is much less than what would be expected from its matched-filter SNR, so the evidence builds up more slowly below $n_{\min}$ (Appendix~\ref{app:other-cases}).
(This behavior is also discussed in Appendix~\ref{app:nstat} in the context of Fig.~\ref{fig:family}.)

\subsection{Information content}
\label{s:information-content}

To understand which mergers carry the information on the population, we use the term $\ell_i = \ln\big[ 1 + \bar N\, ( \bar\Lambda_i - 1 ) \big]$ of segment $i$ in the logarithm of the mixture likelihood~\eqref{eq:mixture2} and its derivative with respect to the seven parameters $\vec\Theta = (\ln R_0, \vec\theta)$,
\begin{equation}
\vec s_i = \frac{\partial \ell_i}{\partial \vec\Theta}\bigg|_{\hat{\vec\Theta}}\,, \qquad
F = \sum_i \vec s_i\, \vec s_i^{\,T}\,.
\label{eq:fisher}
\end{equation}
Here $\hat{\vec\Theta}$ is the maximum of the posterior $p_{\rm I}$ of the catalog-informed search.
The matrix $F$ is the Fisher matrix of the seven parameters in its form with first derivatives.\footnote{This is equivalent in expectation to the second-derivative form of Ref.~\cite{Renzini:2024hiu}.}
Removing segment $i$ moves this maximum by $F^{-1} \vec s_i$, so the product $s_i^{(j)}s_i^{(k)}$ is the contribution of the segment to $F^{jk}$.

\begin{figure*}[!tp]
\centering
\includegraphics[width=\textwidth]{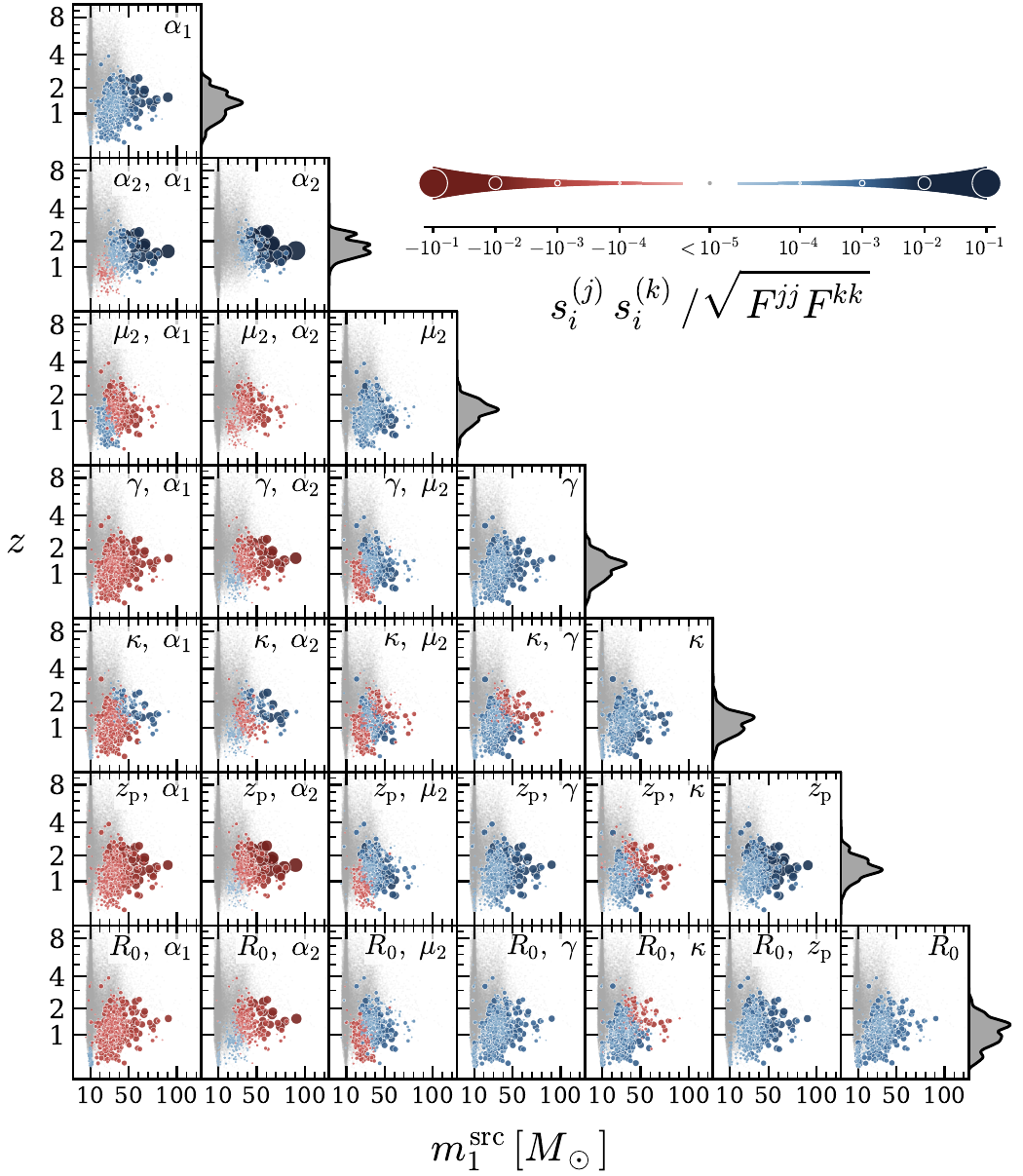}
\caption{Contribution of every unresolved merger of the simulated data to the Fisher matrix $F$ of~\eqref{eq:fisher}, for the source-frame primary mass and the redshift of the merger.
In the panel of parameters $j$ and $k$, the marker size and shade grow with $|s_i^{(j)}\, s_i^{(k)}| / \sqrt{F^{jj} F^{kk}}$.
Markers are blue for positive values and red for negative values.
Mergers with a value below $10^{-5}$ are shown as grey dots.
They reach $m_1^{\rm src} < 10\,M_\odot$ because the analysis bound is on the detector-frame total mass.
The derivatives are computed for $\rho_{\rm th} = 8$ and $n_{\max} = 120$, and the $R_0$ panels refer to $\ln R_0$.
The redshift profile to the right of each diagonal panel shows how the unresolved-merger information on that parameter is distributed in redshift.
Detected mergers and noise-only segments are not shown.}
\label{fig:info_matrix}
\end{figure*}

Figure~\ref{fig:info_matrix} shows which mergers carry this information.
Every panel contains all the unresolved mergers of the simulation at their source-frame primary mass $m_1^{\rm src}$ and redshift $z$.
A merger is drawn as a colored marker with an area proportional to its contribution $s_i^{(j)}s_i^{(k)}$ to the element $F^{jk}$ of the Fisher matrix and as a grey dot when this contribution is negligible.
In every panel, the large markers occupy the same small region of heavy and nearby mergers.
Most of the mergers lie outside this region around the peak of the mass distribution at $10\,M_\odot$ and at $z \simeq 2$, and they are grey in every panel.
Only the nearest mergers of this peak appear as small markers.
These mergers are too quiet to be told apart from noise in their own segment.
At a fixed distance, the SNR of a merger grows with its chirp mass as $\mathcal{M}_c^{5/6}$, so that a binary of the $10\,M_\odot$ peak has about a third of the SNR of a binary of the second peak near $32\,M_\odot$.
At $z \simeq 2$, their likelihood ratio $\bar\Lambda_i$ therefore stays close to one and their derivative $\vec s_i$ close to zero.
These mergers are most of the unresolved mergers and together carry a small part of the information.

The information of the unresolved segments on the population therefore comes from a small fraction of their mergers at the heavy- and nearby-edge of the population.
The panels on the diagonal of Fig.~\ref{fig:info_matrix} show the information on one parameter when the other six are held fixed, and each parameter is constrained by a different part of this region.
The high-mass slope $\alpha_2$ of the mass distribution has the largest markers at the heaviest masses, so a few heavy mergers dominate it.
The location $\mu_2$ of the second peak is constrained by the mergers around the peak itself.
The local rate $R_0$ has many small markers that reach the lowest redshifts, and it takes more from the nearby mergers of the $10\,M_\odot$ peak than any other parameter.

The redshift profiles to the right of the diagonal panels show this information marginalized over the masses.
For every parameter the profile peaks at $z \simeq 1.4$, beyond most of the detected mergers.
The unresolved mergers therefore inform the population at the redshifts where the catalog runs out.
This is why in Fig.~\ref{fig:rz} the catalog fixes the rise of the rate while the unresolved segments bound it at high redshift.

The panels off the diagonal show through the color of a marker whether a merger contributes to the two parameters with the same sign (blue) or with opposite signs (red).
In a panel of a single color, all the mergers contribute with the same sign and the two parameters are degenerate.
This holds among $R_0$, $\gamma$, and $z_p$ because the rate history is normalized at $z = 0$, and it is the degeneracy between the three in Fig.~\ref{fig:pop}.
In the panels containing $\mu_2$, the color changes at the mass of the second peak, so the mergers on the two sides constrain different combinations of the parameters and together separate them.

The detected mergers form the catalog and are not shown in the figure.
Compared with them, the unresolved mergers contribute more to the diagonal elements of $\gamma$, $\kappa$, and $z_p$ and less to those of $R_0$ and the mass parameters.
The noise-only segments are not shown either although together they carry of order ten percent of the information contained in the diagonal elements.

\subsection{Importance of frequency correlations}
\label{s:frequency-correlations}

To confirm that the evidence and the information come from the correlations between frequency bins that enter from the second order on, we analyze a second data set in which these correlations are removed.
In this data set the signal of every merger is replaced by $\tilde h_{I,\nu} \to e^{i\phi_\nu}\, \tilde h_{I,\nu}$ with phases $\phi_\nu$ that are drawn uniformly and independently for each frequency bin and are the same in both detectors.
This replacement keeps the power $|\tilde h_{I,\nu}|^2$ in each detector and the cross spectrum $\tilde h_{1,\nu} \tilde h_{2,\nu}^*$ of every single bin, and it multiplies every product of two different bins by the random phase $e^{i(\phi_\nu - \phi_{\nu'})}$.
We call this data set the {\it phase-randomized} (PR) data. It has the same noise and the same mergers and is analyzed in the same way.

\begin{figure}[!t]
\centering
\includegraphics[width=\columnwidth]{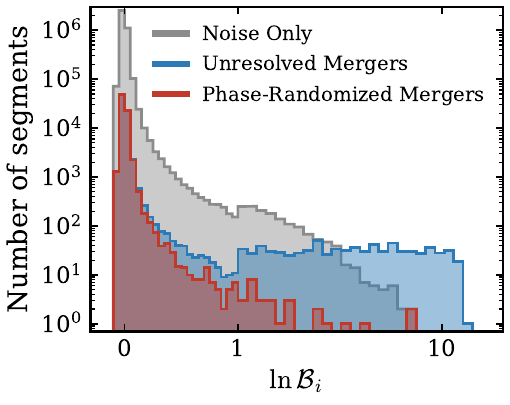}
\caption{Number of noise-only segments and unresolved mergers of the simulated data and of mergers of the PR data in bins of the log Bayes factor $\ln\mathcal{B}_i^{\rm I}$ of the catalog-informed search for $n_{\max} = 120$.
No merger of the PR data is individually detectable.}
\label{fig:twin}
\end{figure}

No segment of the PR data reaches $\rho_{\rm mf}^{\max} = 8$ and the evidence for the background disappears (Fig.~\ref{fig:twin}).
The Bayes factors $\mathcal{B}_i$ of all segments follow the distribution of the noise-only segments, and the uninformed search ends at $\ln\mathcal{B}_{\rm U} = -1.3$ compared to $1.3 \times 10^3$ for the unresolved segments of the simulated data.
\begin{figure}[!t]
\centering
\includegraphics[width=\columnwidth]{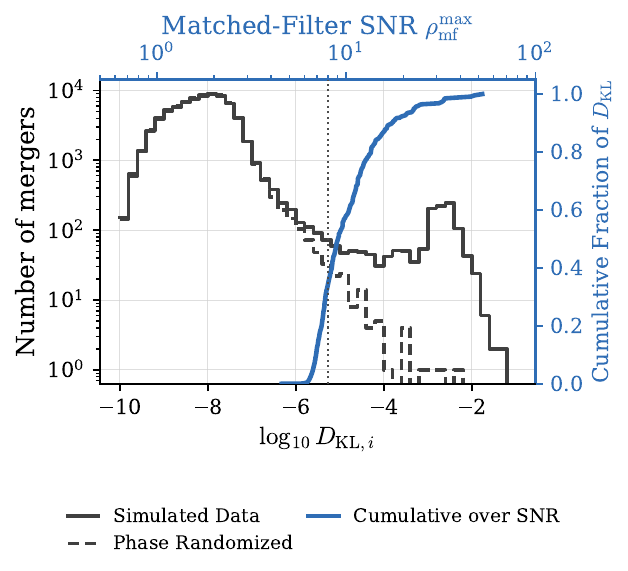}
\caption{Distribution of the information $D_{{\rm KL},i}$ of~\eqref{eq:dkl} over the merger segments for $n_{\max} = 120$, for the simulated data (solid) and the PR data (dashed). The blue curve, read on the top and right axes, is the sum of $D_{{\rm KL},i}$ over the mergers of the simulated data with $\rho^{\max}_{\rm mf}$ below a given value, as a fraction of the sum over all mergers. The dotted line marks the threshold $\rho_{\rm th} = 8$.}
\label{fig:twin_kl}
\end{figure}

The information on the population disappears as well.
We measure the information for segment $i$ with the Kullback--Leibler divergence~\cite{Kullback:1951} between the posteriors with and without the segment~\cite{Johnson:1983}, which to second order in $\vec s_i$ of~\eqref{eq:fisher} is
\begin{equation}
D_{{\rm KL},i} = \frac{1}{2}\, \vec s_i^{\,T} F^{-1}\, \vec s_i\,.
\label{eq:dkl}
\end{equation}
Its sum over all the segments is half the number of parameters, and the unresolved mergers carry about one third of the sum over all of the mergers (Fig.~\ref{fig:twin_kl}).

Figure~\ref{fig:twin_kl} shows the distribution of $D_{{\rm KL},i}$ over the merger segments of the two data sets with $\hat{\vec\Theta}$ and $F$ of the simulated data in both.
In the simulated data, the distribution has a broad bulk at very small values and a separate peak about five orders of magnitude above it, the latter containing the mergers that carry the information.
For the PR data, the bulk is the same and the peak is absent.
The number of mergers of the PR data above the 99.9 percent level of the noise-only segments equals the number expected from noise alone.
Most of the information of the unresolved mergers comes from those with $6 < \rho^{\max}_{\rm mf} \le 8$, for every parameter of the population (Fig.~\ref{fig:twin_kl}).

The results of these analyses demonstrate that
almost all of the evidence and information on the population comes from the correlations between the frequency bins.
A likelihood built only on the statistics of single frequency bins has the same distribution for the two data sets and cannot access them.
\section{Conclusion}
\label{sec:conclusion}

{%
The nearly 400 compact binary mergers, most of them BBHs, that the Advanced LIGO, Advanced Virgo, and KAGRA detectors have detected over the past decade~\cite{LIGOScientific:2026wfs} are the ``tip of the iceberg'' of a much larger population of stellar-mass BBH mergers.
The vast majority of BBH mergers are individually unresolvable at the current detector sensitivity levels, but together they give rise to an intermittent GW background of subthreshold events.
Although this background is known to exist, we have not yet directly detected it with our current searches.

The purpose of this paper was to present a new search for the intermittent background of subthreshold BBH mergers that unifies the standard stochastic-based cross-correlation search and a fully-deterministic template-based search in a single likelihood.
The key quantity of our search is a truncated likelihood ratio, which is {\it fully marginalized} over the extrinsic parameters of the individual mergers.
We represent the data in two detectors by an expansion in terms of approximate sufficient statistics (which are higher-order cross-moments of the data) weighted by geometrical factors that can be interpreted as generalized overlap reduction functions. This expansion is one of the searches of the framework of Ref.~\cite{companion}, and other approximations of the characteristic function give other searches (statistics).
By analyzing one year of simulated data, we demonstrated that our method is {\it fast}, {\it efficient}, and {\it accurate}.
Fast in the sense that it does not require sampling the extrinsic parameters of millions of segments, since their marginalization is computed once in the GORFs (Appendix~\ref{app:construction}).
Efficient in the sense that choosing $n_{\rm max}>32$ allows us to capture all of the relevant information about the subthreshold mergers (Appendix~\ref{app:nstat}, Fig.~\ref{fig:family}).
Accurate in the sense that it recovers the injected rate and the injected parameters of the mass distribution and of the rate history within the 90 percent intervals of the catalog-informed search (Fig.~\ref{fig:pop}).

We also showed the following results in the main text and the appendices.
(i) Once the segments above the threshold $\rho_{\rm th}$ of individual detections are analyzed separately, the terms in the expansion beyond the order $\rho_{\rm th}^2/2$ do not change the Bayes factors or the posteriors of the search.
(ii) A phase-randomized version of the same data shows that almost all of the evidence and of the information on the population come from the {\it correlations between frequency bins}, which a likelihood built on single bins cannot access.
(iii) The mergers below the threshold of individual detections alone still give a clear detection (Sec.~\ref{s:time-to-detection}), and they carry about one third of the information of the mergers on the population (Sec.~\ref{s:frequency-correlations})~\cite{Roulet:2020wyq,Bers:2025tei}.
(iv) The catalog-informed search, which uses information from both the detected and unresolved mergers, is capable of inferring the peak of the redshift distribution of the population of BBH mergers.
All of these results represent a step forward in our understanding of the background of subthreshold mergers.

There is still some work that remains before we can apply this search to real detector data.
(i) First and foremost, our analysis assumes stationary Gaussian noise with a known spectrum.
This is not the case in practice, and the auto moments of the sufficient statistics we use make it sensitive to errors in that spectrum.
Using time slides to measure the noise distribution, or a method like BayesLine~\cite{Littenberg:2014oda} (which both estimates the detector noise and provides an uncertainty level on that estimate) may resolve this problem.
(ii) Second, we need to include non-Gaussian noise transients in the mixture likelihood, since they could otherwise be taken for mergers.
One approach is to extend the mixture likelihood with such transients as a further type of segment with its own likelihood ratio~\cite{Cornish:2014kda,Ashton:2019wvo}.
Then, since a transient in one detector has no counterpart in the other, the cross moments $T^{(n)}_{ab}$ between the two detectors can tell it apart from a merger.
(iii) Third, our GORFs also assume the dominant quadrupole mode with aligned spins.
Higher modes and more detectors add only terms of the same form, with one filtered data and one response for each mode and each detector.

But since none of these limitations changes the structure of the expansion~\eqref{eq:master}, we expect our proposed method to extend to {\it any} search built on the cross-correlation of detector pairs, such as a pulsar timing array where the Hellings--Downs curve~\cite{Hellings:1983fr,Allen:2022ksj} plays the role of the ORF~\cite{Cornish:2013aba}.
We therefore expect a broad range of applications of the search method presented here.
}

\FloatBarrier
\begin{acknowledgments}
Computations presented here used the Minnesota Supercomputing Institute.
We are also grateful to the Neil Cornish, Sharan Banagiri and the LVK Stochastic Group for valuable comments and feedbacks.
TB and VM were in part supported by the NSF grant PHY-2409173.
JDR acknowledges support from NSF grant PHY-2207270 and start-up funds at the University of Texas Rio Grande Valley.
This research has made use of data or software obtained from the Gravitational Wave Open Science Center (gwosc.org), a service of the LIGO Scientific Collaboration, the Virgo Collaboration, and KAGRA. This material is based upon work supported by NSF's LIGO Laboratory which is a major facility fully funded by the National Science Foundation, as well as the Science and Technology Facilities Council (STFC) of the United Kingdom, the Max-Planck-Society (MPS), and the State of Niedersachsen/Germany for support of the construction of Advanced LIGO and construction and operation of the GEO600 detector. Additional support for Advanced LIGO was provided by the Australian Research Council. Virgo is funded, through the European Gravitational Observatory (EGO), by the French Centre National de Recherche Scientifique (CNRS), the Italian Istituto Nazionale di Fisica Nucleare (INFN) and the Dutch Nikhef, with contributions by institutions from Belgium, Germany, Greece, Hungary, Ireland, Japan, Monaco, Poland, Portugal, Spain. KAGRA is supported by Ministry of Education, Culture, Sports, Science and Technology (MEXT), Japan Society for the Promotion of Science (JSPS) in Japan; National Research Foundation (NRF) and Ministry of Science and ICT (MSIT) in Korea; Academia Sinica (AS) and National Science and Technology Council (NSTC) in Taiwan.

We share the companion scripts and example notebooks at \href{https://github.com/dmatter1391/exmarg}{\github}.
\end{acknowledgments}

\FloatBarrier

\onecolumngrid
\appendix

\section*{Appendices}

\section{Alternative derivation of the likelihood ratio}
\label{app:formalism}

In Ref.~\cite{companion}, we wrote the likelihood ratio of a data segment as an average of the characteristic function $z(\eta)$ of the strain over auxiliary variables $\eta$ that stand in for the data.
Expanding $z(\eta)$ in a basis of functions of $\eta$ splits every term into a moment of the strain that depends only on the sources, and an average over $\eta$ that depends only on the data.
Each basis therefore gives its own set of statistics of the data and its own search.
The derivation of Sec.~\ref{sec:method} corresponds to a single basis of this formalism (i.e., powers of $\eta$), and here we recover it from the characteristic function to show that the statistics $T^{(n)}_{ab}$ are one member of this more general family.

By completing the square in the exponent, we can write the Gaussian noise likelihood for each detector $I$ and each frequency bin $\nu$ as a Fourier integral over an auxiliary complex variable $\eta_{I,\nu}$,
\begin{equation}
\frac{1}{\pi \sigma^2_{I,\nu}}\, e^{-|\tilde d_{I,\nu}|^2/\sigma^2_{I,\nu}}
= \int \frac{d^2\eta_{I,\nu}}{4\pi^2}\,
\exp\!\Big[ -\frac{\sigma^2_{I,\nu}}{4} |\eta_{I,\nu}|^2 + i\, \mathrm{Re}\big( \eta_{I,\nu}^*\, \tilde d_{I,\nu} \big) \Big]\,.
\label{eq:fourierLn}
\end{equation}
For signal plus noise, $\tilde d_{I,\nu} - \tilde h_{I,\nu}$ replaces $\tilde d_{I,\nu}$ and the above integrand is multiplied by $e^{-i\,\mathrm{Re}(\eta_{I,\nu}^* \tilde h_{I,\nu})}$.
The average of the product of these factors over the parameters of the source is the characteristic function of the strain
\begin{equation}
z(\eta) \equiv \Big\langle \exp\!\Big[ -\frac{i}{2} \sum_{I,\nu} \big( \eta_{I,\nu}^*\, \tilde h_{I,\nu} + \eta_{I,\nu}\, \tilde h_{I,\nu}^* \big) \Big] \Big\rangle_\pi\,,
\label{eq:charfn}
\end{equation}
and the likelihood ratio becomes $\mathcal{L}_s(d)/\mathcal{L}_n(d) = \langle z(\eta) \rangle_\eta$ with $\langle \cdot \rangle_\eta$ the integral over $\eta$ with the weight of~\eqref{eq:fourierLn} divided by $\mathcal{L}_n(d)$.

We expand $z(\eta)$ in powers of $\eta$, which is the basis that leads to the statistics of Sec.~\ref{sec:method}.
Reference~\cite{companion} keeps any number of mergers per segment through $\langle \exp[\bar N (z - 1)] \rangle_\eta$, and to first order in $\bar N$ this gives the mixture~\eqref{eq:mixture2} with $\mathcal{L}_s(d)/\mathcal{L}_n(d) = \langle z(\eta) \rangle_\eta$, which is the case we expand here.
The waveform~\eqref{eq:template} depends on the coalescence phase only through $e^{-i\varphi_0}$, so the uniform average over $\varphi_0$ keeps only the terms of~\eqref{eq:charfn} with equal powers of its two sums,
\begin{equation}
z(\eta) = \sum_{N=0}^{\infty} \frac{(-1)^N}{4^N (N!)^2}\,
\Big\langle \Big( \sum_{I,\nu} \eta_{I,\nu}^* \tilde h_{I,\nu} \Big)^{\!N}
\Big( \sum_{J,\mu} \eta_{J,\mu} \tilde h_{J,\mu}^* \Big)^{\!N} \Big\rangle_\pi\,.
\label{eq:zeven}
\end{equation}
For fixed source parameters, the $\eta$ integral with the weight of~\eqref{eq:fourierLn} turns each factor of $\eta$ into data.
In one detector and frequency bin,
\begin{equation}
\Big\langle -\frac{i}{2}\, \eta_{I,\nu}^* \Big\rangle_\eta = \frac{\tilde d_{I,\nu}^*}{\sigma^2_{I,\nu}}\,,
\qquad
\Big\langle -\frac{1}{4}\, \eta_{I,\nu}^*\, \eta_{I,\nu} \Big\rangle_\eta = \frac{\tilde d_{I,\nu}^*}{\sigma^2_{I,\nu}}\, \frac{\tilde d_{I,\nu}}{\sigma^2_{I,\nu}} - \frac{1}{\sigma^2_{I,\nu}}\,,
\end{equation}
so each factor becomes the data divided by $\sigma^2_{I,\nu}$, and a pair of $\eta^*$ and $\eta$ in the same bin adds $-1/\sigma^2_{I,\nu}$.
Different bins are integrated independently.
For the $N = 1$ term of~\eqref{eq:zeven}, the factors give $y^2 |\mathcal{Z}|^2$ and the pairs give $-y^2 A$, by~\eqref{eq:zu} and~\eqref{eq:ZA},
\begin{equation}
-\frac{1}{4} \Big\langle \sum_{I,\nu} \eta_{I,\nu}^* \tilde h_{I,\nu} \sum_{J,\mu} \eta_{J,\mu} \tilde h_{J,\mu}^* \Big\rangle_\eta
= y^2 |\mathcal{Z}|^2 - y^2 A\,.
\end{equation}
For the $N = 2$ term, the same rule gives
\begin{equation}
\frac{1}{4}\, y^4 |\mathcal{Z}|^4 - y^4 A\, |\mathcal{Z}|^2 + \frac{1}{2}\, y^4 A^2\,,
\end{equation}
where the three terms have no pair, one pair, and two pairs.
In general, the term of order $N$ is a sum over the number $m$ of pairs,
\begin{equation}
\sum_{m=0}^{N} \frac{1}{(n!)^2\, m!}\, y^{2n} |\mathcal{Z}|^{2n}\, \big( -y^2 A \big)^m\,,
\qquad n = N - m\,.
\label{eq:diagcoef}
\end{equation}
The sum over $m$ at fixed $n$ gives the factor $e^{-y^2 A}$, so that
\begin{equation}
\frac{\mathcal{L}_s(d)}{\mathcal{L}_n(d)} = \sum_{n=0}^{\infty} \frac{1}{(n!)^2} \Big\langle e^{-y^2 A}\, y^{2n} |\mathcal{Z}|^{2n} \Big\rangle_\pi
= \Big\langle e^{-y^2 A}\, I_0\big(2y|\mathcal{Z}|\big) \Big\rangle_\pi\,,
\label{eq:resummed}
\end{equation}
where the factor $e^{-y^2 A}$ is the quadratic term of~\eqref{eq:lnratio} and the Bessel function is the phase integral~\eqref{eq:bessel}.
So, this is the likelihood ratio~\eqref{eq:marg} averaged over the phase.
Separating the data from the geometry as in~\eqref{eq:binomn} and~\eqref{eq:splitorient} then gives the statistics $T^{(n)}_{ab}$ and the GORFs $W^{(n)}_{ab}$ of~\eqref{eq:master}.

\section{Normalizing the likelihood function}
\label{app:norm}

A likelihood ratio has unit mean under the noise distribution, because the signal likelihood in its numerator is normalized,
\begin{equation}
\mathbb{E}_0[\Lambda] = \int \mathcal{D}d\;
\mathcal{L}_n(d)\, \frac{\mathcal{L}_s(d)}{\mathcal{L}_n(d)}
= \int \mathcal{D}d\; \mathcal{L}_s(d) = 1\,.
\label{eq:unitmean}
\end{equation}
But $\Lambda_{n_{\max}}$ does not satisfy this relation.
For a noise-only segment, the derivative of the mixture~\eqref{eq:mixture2} with respect to $\bar N$ has mean $\mathbb{E}_0[\bar\Lambda] - 1$, which vanishes for $\Lambda$ while $\Lambda_{n_{\max}}$ makes it negative, so every noise segment pulls the rate toward zero and the pulls add up over the millions of segments of a search.
Dividing by the sample mean of the noise-only segments does not restore the normalization because this mean is dominated by the rare segments of the tail described in Sec.~\ref{s:foreground-background}.
We therefore compute the normalization analytically.

For fixed extrinsic parameters, the noise enters through $\mathcal{Z} = R_1 \mathcal{Z}_1(t + \tau) + R_2 \mathcal{Z}_2(t)$ of~\eqref{eq:ZA}, which under noise is a complex Gaussian with $\mathbb{E}_0[|\mathcal{Z}|^2] = A$ and the even moments
\begin{equation}
\mathbb{E}_0\big[ |\mathcal{Z}|^{2n} \big] = n!\, A^n\,.
\label{eq:noisemom}
\end{equation}
So, the $n$th term $(y |\mathcal{Z}|)^{2n} / (n!)^2$ of the expansion~\eqref{eq:taylor}, multiplied by the factor $e^{-y^2 A}$ of~\eqref{eq:marg}, has the noise mean
\begin{equation}
\frac{y^{2n}\, n!\, A^n}{(n!)^2}\, e^{-y^2 A}
= e^{-y^2 A}\, \frac{(y^2 A)^n}{n!}\,,
\label{eq:poisweight}
\end{equation}
which is the Poisson weight of mean $y^2 A$
(i.e., a Poisson distribution for $n$ with mean $y^2 A$).
The weights sum to $1$ over all $n$, which recovers~\eqref{eq:unitmean}, and summing them only up to $n_{\max}$ and averaging over the orientation and the distance gives
\begin{equation}
\mathbb{E}_0[\Lambda_{n_{\max}}] = \big\langle
P[\,\mathrm{Pois}(A y^2) \leq n_{\max}\,]
\big\rangle_{\vec\lambda_{\rm ex}, y} = 1 - \epsilon_{n_{\max}}(\vec\lambda_{\rm in})\,.
\label{eq:truncmean}
\end{equation}
This is smaller than $1$ by the averaged Poisson tail
\begin{equation}
\epsilon_{n_{\max}}(\vec\lambda_{\rm in}) = \Big\langle \int_{y_{\min}}^\infty dy\, p(y)\,
P[\,\mathrm{Pois}(A y^2) > n_{\max}\,] \Big\rangle_{\hat n, \psi, \chi}\,,
\label{eq:eps}
\end{equation}
where $P[\mathrm{Pois}(x) > n_{\max}] = \gamma(n_{\max}+1, x)/n_{\max}!$ is the regularized lower incomplete gamma function $P(n_{\max}+1, x)$ of Ref.~\cite{DLMF}, which we tabulate once over the filter powers with the orientation grids of~\eqref{eq:Wdisc}.
The corrected statistic is then
\begin{equation}
\Lambda^\star \equiv \Lambda_{n_{\max}} + \epsilon_{n_{\max}}(\vec\lambda_{\rm in})\,,
\label{eq:lamstar}
\end{equation}
which has $\mathbb{E}_0[\Lambda^\star] = 1$ at every $n_{\max}$, since $\epsilon_{n_{\max}}$ is independent of the data.
The error $\Lambda - \Lambda_{n_{\max}}$ is positive with mean $\epsilon_{n_{\max}}$ under noise, so the correction also cancels the mean error of $\Lambda_{n_{\max}}$ under noise.

In Sec.~\ref{sec:population} we apply the truncated expansion only to the unresolved segments, and the detectable mergers are accounted for by the factor $1 - \bar N(R_0, \vec\theta)\, f_{\rm det}(\vec\theta)$ of~\eqref{eq:LU}.
The term added to $\Lambda_{n_{\max}}(M^{\rm det}, z)$ in~\eqref{eq:lamstar} therefore becomes
\begin{equation}
\epsilon_{n_{\max}}(M^{\rm det}, z) - P\big(\rho_{\rm mf}^{\max} > \rho_{\rm th} \mid M^{\rm det}, z\big)
\end{equation}
where this difference is positive and zero elsewhere.
Here $\epsilon_{n_{\max}}(M^{\rm det}, z)$ is~\eqref{eq:eps} at the fixed distance $y = D_0/D(z)$ of Sec.~\ref{s:marg-external} without the integral over $y$, and $P(\rho_{\rm mf}^{\max} > \rho_{\rm th} \mid M^{\rm det}, z)$ is the probability of~\eqref{eq:alpha} that a merger of mass $M^{\rm det}$ at redshift $z$ is detected.
The Poisson distribution in~\eqref{eq:eps} has the mean $A y^2 = \rho_{\rm opt}^2/2$ of~\eqref{eq:snr}, so for $n_{\max} > \rho_{\rm th}^2/2$ the tail $\epsilon_{n_{\max}}(M^{\rm det}, z)$ comes from mergers with $\rho_{\rm opt} > \rho_{\rm th}$ that are detected.
The added term then vanishes and the results of Sec.~\ref{sec:population} for $\rho_{\rm th} = 8$ and $n_{\max} = 40$, 80 and 120 contain no correction.
We use $\Lambda^\star$ in the comparison with nested sampling of Appendix~\ref{app:BF-convergence}, where single segments louder than $\rho_{\rm mf} = \sqrt{2 n_{\max}}$ enter the truncated expansion, and in the distributions of Fig.~\ref{fig:bf_orders}.

\begin{figure*}[!htbp]
\centering
\includegraphics[width=0.85\textwidth]{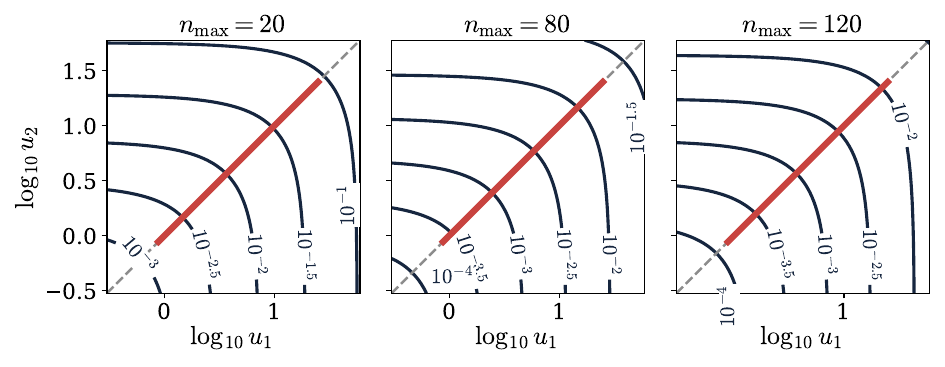}
\caption{The normalization $\epsilon_{n_{\max}}(u_1, u_2)$ of~\eqref{eq:eps} in the plane of the two filter powers for three values of $n_{\max}$, under the Madau--Dickinson distance prior of Fig.~\ref{fig:tables}.
It depends on the filter powers only through $A = u_1 |R_1|^2 + u_2 |R_2|^2$.
The dashed line is the diagonal $u_1 = u_2$, and the red segment is the range of filter powers of detector-frame total masses between 20 and $200\,M_\odot$.}
\label{fig:eps}
\end{figure*}

Figure~\ref{fig:eps} shows $\epsilon_{n_{\max}}$ in the plane of the two filter powers for the Madau--Dickinson distance prior of Fig.~\ref{fig:tables}, where the analysis lies on the diagonal $u_1 = u_2 \equiv u$.
Along the red segment of Fig.~\ref{fig:eps}, $\epsilon_{n_{\max}}$ grows with $u$, because a larger filter power makes $y^2 A$ exceed $n_{\max}$ out to a larger distance, so a larger part of the distance prior falls in the tail of~\eqref{eq:eps}.

\section{Pre-computing the geometrical factors}
\label{app:construction}

This appendix describes how we compute the GORFs once on fixed grids of the sky position and the orientation, and it gives their resolution and their cost.
We start by evaluating the distance moment~\eqref{eq:Mdist} for the uniform-in-volume prior of Sec.~\ref{sec:method}, where the substitution $t = y^2 A$ gives the upper incomplete gamma function~\cite{DLMF},
\begin{equation}
M_n(A) = \frac{3}{2}\, y_{\min}^3\, A^{\frac{3-2n}{2}}\,
\Gamma\!\left( \frac{2n-3}{2},\, y_{\min}^2 A \right).
\label{eq:Mn}
\end{equation}
The bare moments $\int dy\, p(y)\, y^{2n}$ of the prior diverge for every $n \geq 2$, since arbitrarily close sources have arbitrarily large amplitudes.
The moment $M_n(A)$ is finite because the quadratic term $e^{-y^2 A}$ of~\eqref{eq:marg} cuts off the loud tail at $y \simeq 1/\sqrt{A}$.
For the Madau--Dickinson distance prior of Fig.~\ref{fig:tables}, we evaluate $M_n(A)$ numerically.

\begin{figure}[!htbp]
\centering
\includegraphics[width=0.49\textwidth]{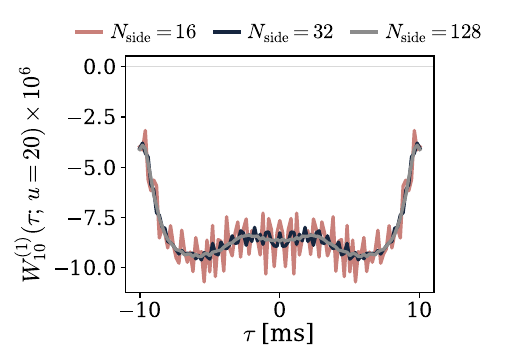}
\caption{The first-order GORF $W^{(1)}_{10}(\tau)$ evaluated at the filter power $u = 20$ under the Madau--Dickinson distance prior used for Fig.~\ref{fig:tables}.
We use HEALPix grids with $N_{\rm side} = 16$, 32, and 128.}
\label{fig:Wnside}
\end{figure}

With the sky pixelated by a HEALPix grid of $N_{\rm pix}$ pixels, the polarization on a uniform grid of $n_\psi$ angles, and the inclination angle, the GORF of~\eqref{eq:W} is
\begin{equation}
W^{(n)}_{ab}(\tau_m, u_1, u_2) = \frac{1}{N_{\rm pix}\, n_\psi}
\sum_{\rm pix} \sum_{\psi} \sum_{\chi} \frac{w_\chi}{2}\,
M_n(A)\, R_1^a R_1^{*b} R_2^{n-a} R_2^{*(n-b)}\,
\Delta_m\!\big(\hat n \cdot \Delta \vec x / c\big)\,.
\label{eq:Wdisc}
\end{equation}
Here, the Dirac delta of~\eqref{eq:W} becomes
\begin{equation}
\Delta_m(\tau) \equiv \max\!\big( 0,\, 1 - |\tau - \tau_m| / \delta\tau \big)\,,
\label{eq:deposit}
\end{equation}
which spreads the delay of each pixel linearly over the two neighboring points of a uniform delay grid of spacing $\delta\tau$, spanning the light-travel time between the two detectors.
Since the two weights of a pixel add up to one, $W^{(n)}_{ab}(\tau_m, u_1, u_2)$ of~\eqref{eq:Wdisc} carries the integral of~\eqref{eq:W} over the cell of width $\delta\tau$ around $\tau_m$, and the integrals over $\tau$ in~\eqref{eq:master} and~\eqref{eq:read} are sums over the delay grid.
The sky resolution sets the accuracy of the GORFs, since the delay distribution inherits the pixel scale.
Figure~\ref{fig:Wnside} shows that for $N_{\rm side} = 16$, the GORF oscillates at the pixel scale, while for $N_{\rm side} = 32$, the oscillation shrinks to a ripple around the $N_{\rm side} = 128$ curve.
Appendix~\ref{app:ns-agreement} tests the resolution $N_{\rm side} = 32$ of our tables against a nested sampler that integrates the orientation numerically.

We precompute the GORFs at a fixed distance, where they depend on the filter power and the distance only through $s = u y^2$, so we tabulate them on a grid of $s$. In the analysis of Sec.~\ref{sec:population}, a merger is described by its detector-frame total mass $M^{\rm det}$ and its redshift $z$. For each mass, the redshift grid is uniform in the logarithm of the optimal SNR $\rho_{\rm opt}$ of~\eqref{eq:snr} up to $z = 10$, which gives about $1.2 \times 10^4$ grid points in total.

We also precompute the sum over the orders of the series, exchanging the time average of~\eqref{eq:T} with the sum over orders.
At a given delay, each order of~\eqref{eq:master} depends on the data only through the two magnitudes $|\mathcal{Z}_1|$ and $|\mathcal{Z}_2|$, and the relative phase $\Delta\varphi \equiv \arg( \mathcal{Z}_1 \mathcal{Z}_2^* )$,
\begin{equation}
\begin{aligned}
\mathcal{S}^{(n)}(\tau, \mathcal{Z}_1, \mathcal{Z}_2; u_1, u_2) = {} & \frac{1}{(n!)^2}
\sum_{a=0}^{n} \sum_{b=0}^{n} \binom{n}{a} \binom{n}{b}W^{(n)}_{ab}(\tau, u_1, u_2)\,
|\mathcal{Z}_1|^{a+b}\, |\mathcal{Z}_2|^{2n-a-b}\cos[(a-b)\, \Delta\varphi]\,.
\end{aligned}
\label{eq:kernelS}
\end{equation}
But these two magnitudes do not enter independently.
By writing $|\mathcal{Z}_1|^{a+b} |\mathcal{Z}_2|^{2n-a-b} = |\mathcal{Z}_2|^{2n} e^{(a+b)\kappa}$, where $\kappa \equiv \ln |\mathcal{Z}_1| - \ln |\mathcal{Z}_2|$ is the log ratio of the two magnitudes,
we are able to separate
each order into a power of $|\mathcal{Z}_2|$ and a function of the log ratio alone,
\begin{equation}
\begin{aligned}
&\mathcal{S}^{(n)}(\tau, \mathcal{Z}_1, \mathcal{Z}_2; u_1, u_2)
= |\mathcal{Z}_2|^{2n}\,
g_n(\tau, \kappa, \Delta\varphi; u_1, u_2)\,,
\quad{\rm where}
\\
&g_n(\tau, \kappa, \Delta\varphi; u_1, u_2)
\equiv \frac{1}{(n!)^2} \sum_{a=0}^{n} \sum_{b=0}^{n}
\binom{n}{a}\binom{n}{b}W^{(n)}_{ab}(\tau, u_1, u_2)\,
e^{(a+b)\kappa}\cos[(a-b)\,\Delta\varphi]\,.
\end{aligned}
\label{eq:kernelg}
\end{equation}
Summing over the orders then gives a single function of four data arguments,
\begin{equation}
\begin{aligned}
&\mathcal{S}(\tau, \kappa, \Delta\varphi, |\mathcal{Z}_2|; u_1, u_2)\equiv \sum_{n \le n_{\max}} |\mathcal{Z}_2|^{2n}\,
g_n(\tau, \kappa, \Delta\varphi; u_1, u_2)
\end{aligned}
\label{eq:kernelsum}
\end{equation}
which we tabulate at every point of the grid of filter powers.
The cost of analyzing a segment is the same at every $n_{\max}$, and we carry the GORFs to $n_{\max} = 120$, beyond the minimum order $n_{\min}$ of Sec.~\ref{s:searches} above which the Bayes factors and the posteriors no longer change.

When the data of a segment arrive, each set of intrinsic parameters $\vec\lambda_{\rm in}$ needs only its filtered data $\mathcal{Z}_I(t, \vec\lambda_{\rm in})$ and its filter powers $u_I(\vec\lambda_{\rm in})$ of~\eqref{eq:zu}.
We read the table of~\eqref{eq:kernelsum} at these values and assemble the marginalized likelihood ratio as
\begin{equation}
\begin{aligned}
\Lambda(\vec\lambda_{\rm in}) = \frac{1}{2\Delta t} \int dt \int d\tau\,
\mathcal{S}\big(&\tau,\, \kappa(t, \tau, \vec\lambda_{\rm in}),\,
\Delta\varphi(t, \tau, \vec\lambda_{\rm in}), |\mathcal{Z}_2(t, \vec\lambda_{\rm in})|;\,
u_1(\vec\lambda_{\rm in}),\, u_2(\vec\lambda_{\rm in}) \big)\,,
\end{aligned}
\label{eq:read}
\end{equation}
where the data arguments $\kappa$ and $\Delta\varphi$ are given by the two filtered data at the delay $\tau$,
\begin{equation}
\begin{aligned}
\kappa(t, \tau, \vec\lambda_{\rm in}) &= \ln |\mathcal{Z}_1(t + \tau, \vec\lambda_{\rm in})|
- \ln |\mathcal{Z}_2(t, \vec\lambda_{\rm in})|\,, \\
\Delta\varphi(t, \tau, \vec\lambda_{\rm in}) &=
\arg\!\big[ \mathcal{Z}_1(t + \tau, \vec\lambda_{\rm in})\,
\mathcal{Z}_2^*(t, \vec\lambda_{\rm in}) \big]\,.
\end{aligned}
\label{eq:readargs}
\end{equation}

The tables for the 53 grid points of the filter power occupy 40 GB.
After this setup, the cost of analyzing a segment is the same for $n_{\max} = 1$, 5, and 120.
Analyzing one segment for 120 masses takes about 0.1~s on one NVIDIA L40S GPU.
So analyzing a year of data on the grid above takes about a hundred GPU hours.
For comparison, a nested-sampling evaluation of the same likelihood ratio takes about an hour on one CPU core for a single segment and template (Appendix~\ref{app:BF-convergence}).

\section{Stochastic-to-deterministic transition}
\label{app:nstat}

Equation~\eqref{eq:master} is exact but in practice we truncate the expansion over orders at $n_{\max}$.
In this appendix, we ask how many orders a signal needs as its SNR grows, and whether $\Lambda_{n_{\max}}$ gives accurate likelihood ratios and posteriors.

\subsection{Likelihood ratio vs $n_{\rm max}$ for single events}
\label{s:single}

\begin{figure*}[!htbp]
\centering
\includegraphics[width=0.5\textwidth]{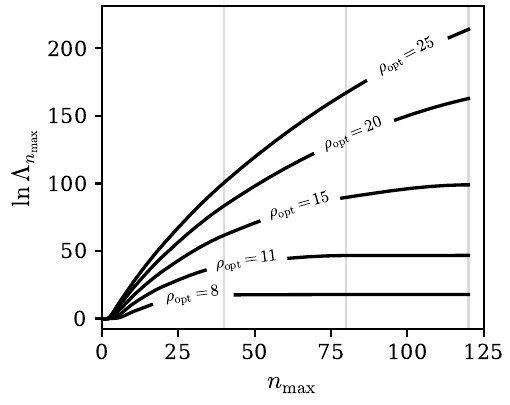}
\caption{The passage from the stochastic to the deterministic description of the data for a single nonspinning binary.
The curves show $\ln \Lambda_{n_{\max}}$ at the injected parameters versus $n_{\max}$ for the binary at five distances, labeled by their $\rho_{\rm opt}$, calculated using the O4 noise curve of Sec.~\ref{sec:population}.
The detector frame parameters stay at the injected values of Fig.~\ref{fig:family_post}.
The vertical grey lines mark the orders $n_{\max} = 40$, 80 and 120 of Fig.~\ref{fig:family_post}.
The analysis marginalizes over the extrinsic parameters $\vec\lambda_{\rm ex}$ of~\eqref{eq:ls}, injected at polar and azimuthal sky angles $(2.54, 1.30)~{\rm radians}$ and $(t, \varphi_0, \psi, \chi) = (2.41~{\rm s}, 2.23, 0.07, 0.98)$.}
\label{fig:family}
\end{figure*}

We start with a single nonspinning binary in a short segment of simulated data for the LIGO Hanford and LIGO Livingston detectors and move it closer to Earth while the noise realization stays fixed.
We use $\Lambda_{n_{\max}}$ as the likelihood of the posterior, which we evaluate on a grid over the three intrinsic parameters $(\Mc, q, \chi_{\rm eff})$ of the aligned spin IMRPhenomXAS waveform~\cite{XAS}, with the binary injected at $\chi_{\rm eff} = 0$.
Figure~\ref{fig:family} shows $\ln \Lambda_{n_{\max}}$ of the event against $n_{\max}$ at five luminosity distances, labeled by the optimal SNR $\rho_{\rm opt}$ of~\eqref{eq:snr}.
Each curve rises with $n_{\max}$ and levels off at an order that grows as the source comes closer, with the two loudest sources still rising at $n_{\max} = 120$.
A distant source at small SNR is therefore described by the first few orders, which stay close to the stochastic description, while a nearby source at large SNR needs many orders before its likelihood ratio reaches that of its full waveform.

\begin{figure}[!htbp]
\centering
\includegraphics[width=0.5\columnwidth]{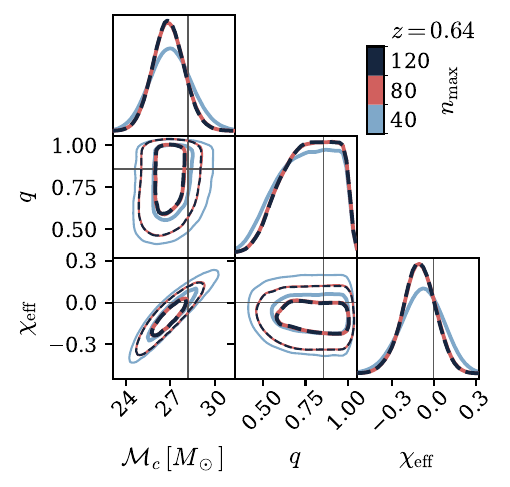}
\caption{Posterior of the binary of Fig.~\ref{fig:family} with the O4 noise curve over the intrinsic parameters $\vec\lambda_{\rm in} = (\Mc, q, \chi_{\rm eff})$ at $\rho_{\rm opt} = 11$ (labeled $z = 0.64$ in the figure) for $n_{\max}=40$, 80, and 120, with 50 and 90 percent credible regions and grey lines at the injected values $(28.2\,M_\odot, 0.86, 0)$.
The grid spans the intrinsic parameter priors of Appendix~\ref{app:ns-agreement}, uniform in the detector frame total mass in $[20, 200]\,M_\odot$, $q \in [0.3, 1]$ and $\chi_{\rm eff} \in [-0.6, 0.6]$, and the axes span the range of the posterior samples.
Note that $\Mc = m_1^{\rm det}\,q^{3/5}/(1+q)^{1/5}$ is the detector frame chirp mass, constructed from the detector frame primary mass $m_1^{\rm det}$ and the mass ratio $q \le 1$.}
\label{fig:family_post}
\end{figure}

Figure~\ref{fig:family_post} shows the same passage in the posterior of the event at $\rho_{\rm opt} = 11$, which is slightly wider at $n_{\max} = 40$ than at $n_{\max} = 80$ and 120, where it no longer changes.
The posterior is broad at low order because $\Lambda_{n_{\max}}$ grows at most as $|\mathcal{Z}|^{2 n_{\max}}$ and therefore rises toward its peak more slowly than $\Lambda$.
The saturation follows the condition $n_{\max} > \rho_{\rm mf}^2/2$ of Sec.~\ref{sec:method}, which comes from the distance integral.
The distance integrand $y^{2n} e^{-y^2 A}$ of the $n$th term of~\eqref{eq:taylor} peaks at $y^2 = n/A$.
At this peak, the term is proportional to $(|\mathcal{Z}|^2/A)^n/n!$, up to the prior and a power of $n$, which is a Poisson weight with mean $|\mathcal{Z}|^2/A = \rho_{\rm mf}^2/2$.
So the terms grow up to $n \approx \rho_{\rm mf}^2/2$ and fall off beyond it.
Appendix~\ref{app:BF-convergence} measures the convergence against the injected $\rho_{\rm opt}$ of Eq.~\eqref{eq:snr}.

\subsection{Convergence of the likelihood ratio}
\label{app:BF-convergence}

\begin{figure}[!htbp]
\centering
\includegraphics[width=\textwidth]{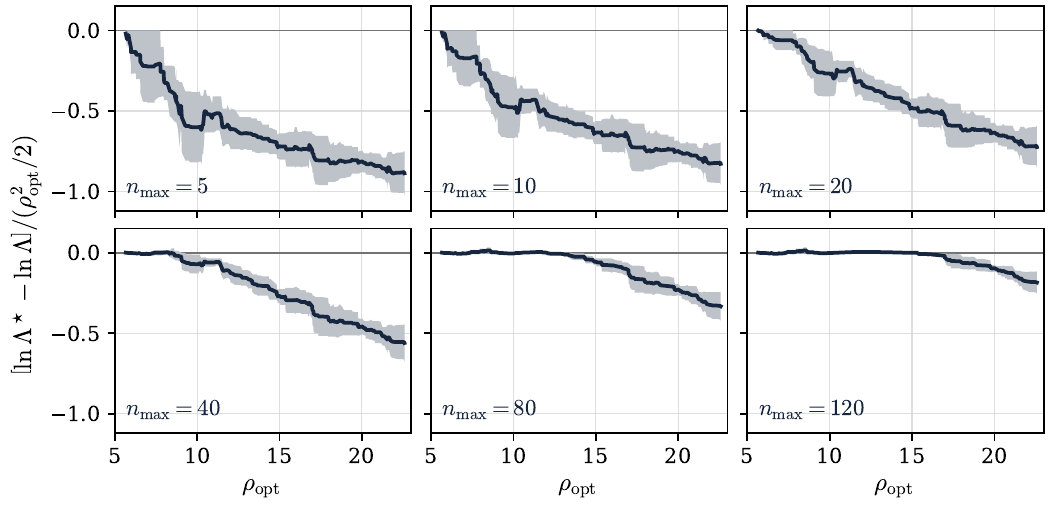}
\caption{$\Lambda^\star$ of~\eqref{eq:lamstar} against an independent reference $\Lambda$, computed from nested sampling, both evaluated at the injected intrinsic parameters of 252 events, and for six values of $n_{\max}$.
The curve is the median and the band the 16th to 84th percentile of the difference of the two logarithms divided by $\rho_{\rm opt}^2/2$.}
\label{fig:reference}
\end{figure}

We test the above condition on the truncation order $n_{\max}$ against an independent evaluation of $\Lambda$ by nested sampling~\cite{dynesty} over the extrinsic parameters, on events that span the SNRs of a background.
This reference has a statistical error of its own that we keep well below the differences produced by the truncation, so it serves as a controlled test of the corrected statistic $\Lambda^\star$ of~\eqref{eq:lamstar}.
Each reference evaluation takes about an hour on one CPU core for a single segment and template.
The two logarithms agree up to an SNR that grows with $n_{\max}$ as this condition predicts, as illustrated in Fig.~\ref{fig:reference}.

For the comparison in Fig.~\ref{fig:reference}, the nested sampler uses the phase integral~\eqref{eq:bessel} in closed form and integrates over the distance with the same quadrature as our analysis, so that the two differ only by the truncation of the series.
The events are equal mass nonspinning binaries drawn from the GWTC-5 mass model with a Madau--Dickinson rate, and both sides use the same data and priors, and both are evaluated at the injected intrinsic parameters.
We take 21 events in each of twelve bins of the optimal SNR $\rho_{\rm opt}$ of \eqref{eq:snr} between 5.7 and 22.6.
For a loud merger, $\ln\Lambda$ is close to $\rho_{\rm opt}^2/2$, so the difference of the two logarithms divided by $\rho_{\rm opt}^2/2$ is the fraction of $\ln\Lambda$ lost to the truncation. It vanishes where the series has converged and falls toward $-1$ where $\ln\Lambda_{n_{\max}}$ stays far below $\rho_{\rm opt}^2/2$.

\subsection{Recovery of event parameters}
\label{app:unbiased-pop-parameters}

\begin{figure*}[!htbp]
\centering
\includegraphics[width=0.42\textwidth]{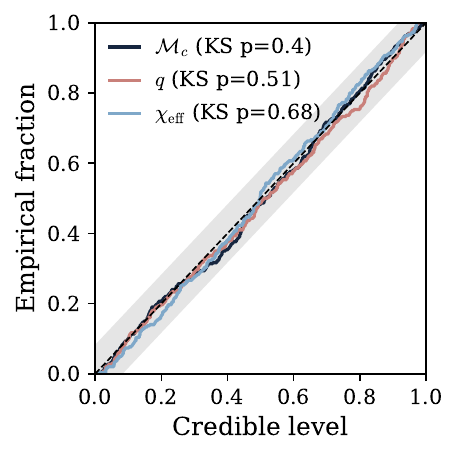}\hspace{0.7in}
\includegraphics[width=0.42\textwidth]{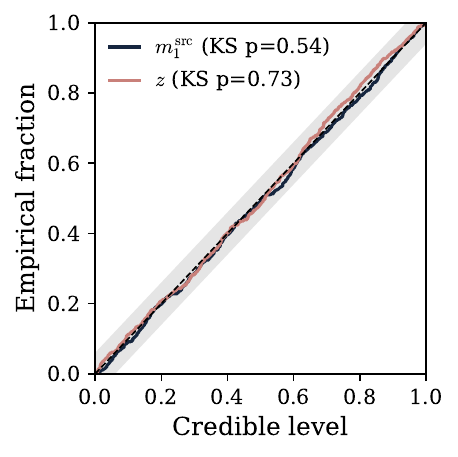}
\caption{PP curves for $n_{\max} = 120$. The horizontal axis is the credible level, and the vertical axis is the fraction of injections whose true value lies within the credible interval of that level.
Left, the three intrinsic parameters $(\Mc, q, \chi_{\rm eff})$ of Fig.~\ref{fig:family_post} for 300 injections.
Right, $(m_1^{\rm src}, z)$ for 592 injections.
The band is the 95 percent Kolmogorov--Smirnov (KS) region.}
\label{fig:pp}
\end{figure*}

The probability--probability (PP) curves of Fig.~\ref{fig:pp} for injections drawn from the priors are consistent with the diagonal. This demonstrates that the analysis recovers the parameters of single events without bias when the priors used for the simulation and for the analysis are the same.

To produce the left panel of Fig.~\ref{fig:pp}, we analyzed 300 injections drawn from the analysis priors, each in its own noise realization and at $n_{\max} = 120$.
For the right panel, we kept the source frame primary mass and the redshift of an equal-mass nonspinning binary as parameters, with GORFs at a fixed distance in which the distance moment $M_n(A)$ of~\eqref{eq:Mdist} is replaced by $y^{2n} e^{-y^2 A}$.
We then analyzed 592 injections drawn from the GWTC-5 population, which is also the prior of the analysis (Appendix~\ref{app:pop-priors}).
These GORFs carry $y = D_0/D(z)$ as a further argument, so the redshift enters the likelihood ratio~\eqref{eq:master} through $y$ and through the detector frame mass $M^{\rm det} = 2 m_1^{\rm src}(1 + z)$ of the filtered data~\eqref{eq:zu}.

\subsection{Agreement with nested sampling}
\label{app:ns-agreement}

\begin{figure*}[!htbp]
\centering
\includegraphics[width=0.9\textwidth]{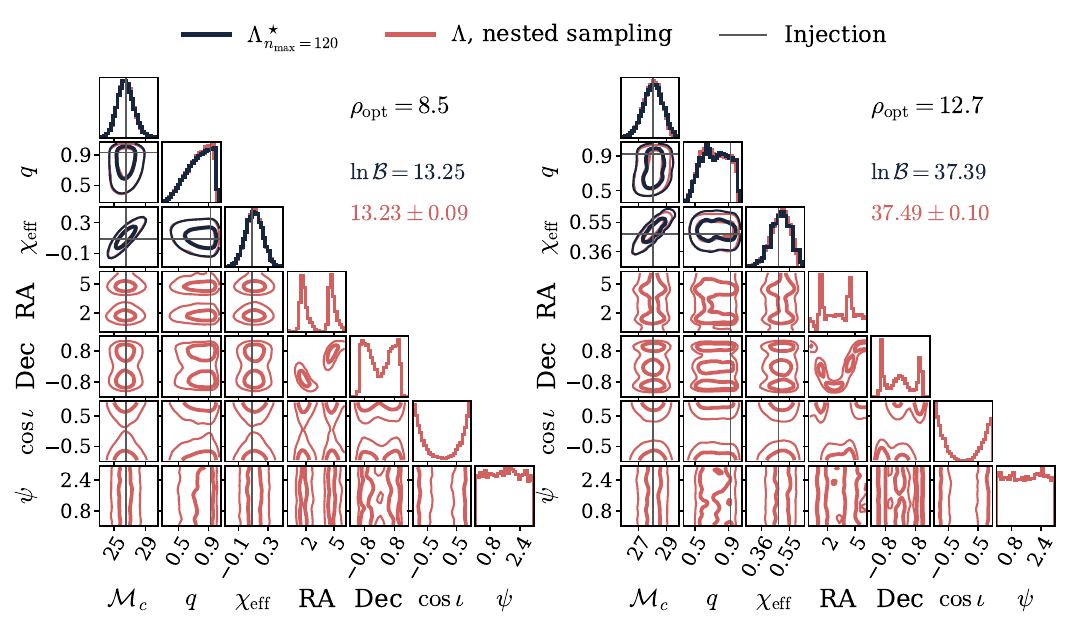}
\caption{Posteriors of two injected aligned spin binaries from our analysis for $n_{\max} = 120$ and from the nested sampler for the detector-frame chirp mass $\Mc$ (of Fig.~\ref{fig:family_post}), the mass ratio $q$, and effective spin $\chi_{\rm eff}$.
The upper-right corner of each block gives the SNR $\rho_{\rm opt}$ and log Bayes factor (relative to noise) for our calculation and nested sampling (including the sampling error of the sampler), where the Bayes factor of our calculation is the integral of $\Lambda_{n_{\max}}$ over the priors on $(\Mc, q, \chi_{\rm eff})$.
The panels for right ascension (RA), declination (Dec), $\cos\iota$ and $\psi$ apply only to the sampler, and the grey lines mark the injected values of $(\Mc, q, \chi_{\rm eff})$.
Both analyses use the priors of Appendix~\ref{app:ns-agreement}, uniform in the detector frame total mass and in volume out to $D = 82$~Gpc.}
\label{fig:ns}
\end{figure*}

\begin{figure*}[!htbp]
\centering
\includegraphics[width=0.9\textwidth]{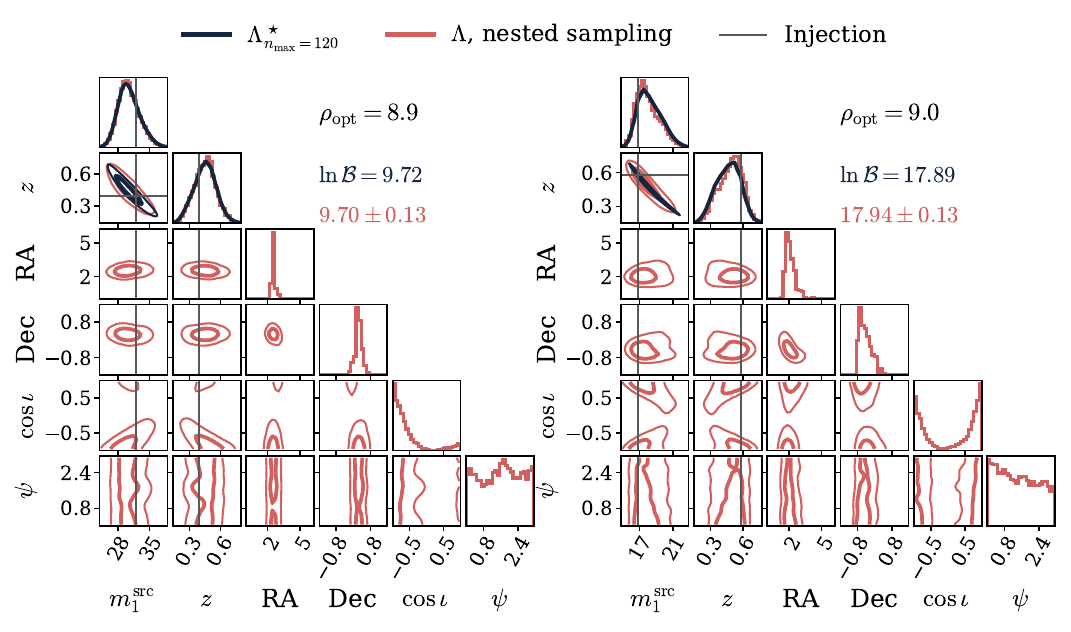}
\caption{Same as Fig.~\ref{fig:ns}, but for the source-frame primary mass $m_1^{\rm src}$ and the redshift $z$ as the parameters.}
\label{fig:ns2p}
\end{figure*}

The posterior distributions and Bayes factors against noise of individual events agree with those of the nested sampler as shown in Figs.~\ref{fig:ns} and \ref{fig:ns2p}.
The nested sampler explores the sky position, inclination, and polarization of the source numerically, while our analysis integrates them once on the grids of~\eqref{eq:Wdisc} inside $W^{(n)}_{ab}$.
So, the agreement that we find also verifies that we precomputed the GORFs with high-enough
resolution that their discretization did not change the likelihood evaluation, as discussed in Appendix~\ref{app:construction}.

For Fig.~\ref{fig:ns}, we injected two aligned spin binaries into their own segments of simulated noise and ran the nested sampler \texttt{dynesty}~\cite{dynesty} in \texttt{bilby}~\cite{bilby} with 2000 live points on the same two segments and with the same waveform and priors.
The priors were uniform in the total detector frame mass in $[20, 200]\,M_\odot$, $q\in [0.3, 1]$, $\chi_{\rm eff}\in[-0.6, 0.6]$, and volume out to $D = 82$~Gpc.
For Fig.~\ref{fig:ns2p}, we used the source frame primary mass and the redshift on twelve events with $\rho_{\rm opt}$ between 7.3 and 9.0, under a prior uniform in the total detector frame mass $M^{\rm det} = 2 m_1^{\rm src} (1 + z)$ in $[10, 500]\,M_\odot$ and in volume out to $D = 82$~Gpc, which is the density $(1 + z)\, D^2\, dD/dz$ in the fitted parameters.
The Bayes factors agree for all twelve events, and Fig.~\ref{fig:ns2p} shows two of them.
Together the right panel of Fig.~\ref{fig:pp} and Fig.~\ref{fig:ns2p} show that the redshift of a merger is recovered from the same statistics $T^{(n)}_{ab}$ with the fixed distance GORFs of Appendix~\ref{app:unbiased-pop-parameters}. On the same data, $\Lambda^\star$ and its posteriors also agree with the importance sampling of \texttt{cogwheel}~\cite{Roulet:2024hwz}, which has a sampling scatter and needs several seconds per segment and template on one CPU core, while $\Lambda^\star$ has no scatter and takes less than 1~s (see the notebooks of the companion repository).

\section{Population model and priors}
\label{app:pop-priors}

This appendix defines the population model with its seven parameters and the uninformed prior $\pi_0$.
The distribution $\pi_{\rm src}(m_1^{\rm src} \mid \vec\theta)$ of the source-frame primary mass is the GWTC-5 model~\cite{GWTC5} with the low-mass and high-mass slopes $\alpha_1$ and $\alpha_2$ of its broken power law and the location $\mu_2$ of its second peak as free parameters, and with its other hyperparameters held at their injected values.
The merger rate density is $R(z) = R_0\, \psi(z \mid \vec\theta)$ with the Madau--Dickinson form~\cite{MadauDickinson2014}
\begin{equation}
\psi(z \mid \vec\theta) = \frac{(1+z)^{\gamma}\, \big[ 1 + (1+z_p)^{-\kappa} \big]}{1 + \big[ (1+z)/(1+z_p) \big]^{\kappa}}\,,
\label{eq:rz}
\end{equation}
and $\vec\theta \equiv (\alpha_1, \alpha_2, \mu_2, \gamma, \kappa, z_p)$.
The injected values are $R_0 = 19.1~{\rm Gpc^{-3}\,yr^{-1}}$, $\alpha_1 = 1.48$, $\alpha_2 = 5.42$, and $\mu_2 = 32.3\,M_\odot$, which are the median values of the GWTC-5 analysis~\cite{GWTC5}, and $\gamma = 2.7$, $\kappa = 5.6$, and $z_p = 1.9$ of Ref.~\cite{MadauDickinson2014}.
The uninformed prior $\pi_0$ assigns the same probability density to every point $(\ln R_0, \alpha_1, \alpha_2, \mu_2, \gamma, \kappa, z_p)$ inside the ranges
\begin{equation}
\begin{aligned}
R_0/({\rm Gpc^{-3}\,yr^{-1}}) &\in [1, 300]\,, & \alpha_1 &\in [-2, 6]\,, & \alpha_2 &\in [1, 10]\,, & \mu_2/M_\odot &\in [20, 50]\,, \\
\gamma &\in [-2, 10]\,, & \kappa &\in [-2, 10]\,, & z_p &\in [0.5, 4]\,,
\end{aligned}
\label{eq:pi0}
\end{equation}
and zero probability outside them.
So it is uniform in $\ln R_0$ and in each of the six parameters of $\vec\theta$.

For a population $(R_0, \vec\theta)$, the number of mergers in a segment per unit detector-frame total mass $M^{\rm det} = 2 m_1^{\rm src} (1 + z)$ and redshift $z$ is
\begin{equation}
\frac{d\bar N}{dM^{\rm det}\, dz} = R_0\, T_{\rm seg}\, \frac{dV_c}{dz}\, \frac{\psi(z \mid \vec\theta)}{2(1+z)^2}\; \pi_{\rm src}\Big( \frac{M^{\rm det}}{2(1+z)} \,\Big|\, \vec\theta \Big)\,,
\label{eq:dnbar}
\end{equation}
and its integral over the redshifts and the masses of the analysis gives the mean number of mergers in a segment and the distribution of their mass and redshift,
\begin{align}
\bar N(R_0, \vec\theta) &= \int_0^{10} dz \int_{20\,M_\odot}^{1000\,M_\odot} dM^{\rm det}\; \frac{d\bar N}{dM^{\rm det}\, dz}\,,
\label{eq:nbarpop} \\
\pi(M^{\rm det}, z \mid \vec\theta) &= \frac{1}{\bar N(R_0, \vec\theta)}\, \frac{d\bar N}{dM^{\rm det}\, dz}\,.
\label{eq:pidet}
\end{align}
The average~\eqref{eq:lambar} is evaluated on a grid of masses and redshifts (Appendix~\ref{app:construction}).
The probability $P(\rho_{\rm mf}^{\max} > \rho_{\rm th} \mid M^{\rm det}, z)$ that a merger of mass $M^{\rm det}$ at redshift $z$ is detected enters the detected fraction $f_{\rm det}(\vec\theta)$ of~\eqref{eq:alpha}, and we measure it with about $10^7$ injections into simulated noise.
For the thresholds that we use, no segment with noise alone exceeds $\rho_{\rm th}$.

\section{Truncation order and threshold values}
\label{app:other-cases}

\begin{figure*}[!htbp]
\centering
\includegraphics[width=\textwidth]{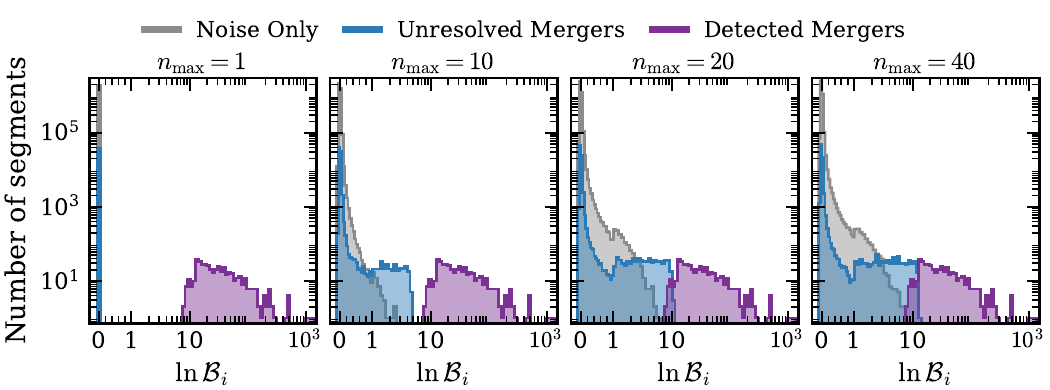}
\caption{Same as Fig.~\ref{fig:year_lnb} for $n_{\max} = 1$, 10, 20, and 40 with the truncated series normalized as in Appendix~\ref{app:norm}.
The detected segments use the exact likelihood ratio in every panel.}
\label{fig:bf_orders}
\end{figure*}

To test the minimum order $n_{\min} = \rho_{\rm th}^2/2$, we repeat the catalog-informed and uninformed searches  of
Secs.~\ref{s:searches} and \ref{s:popparms-main} for four values of $n_{\max}$, and for $\rho_{\rm th} = 8$ and $12$, and we add the analysis with no threshold in which every segment is truncated at $n_{\max}$.
The orders below $n_{\min}$ are $n_{\max} = 20$ for $\rho_{\rm th} = 8$, $n_{\max} = 20$ and 40 for $\rho_{\rm th} = 12$, and every order of the analysis with no threshold.
For these orders, the posteriors and the Bayes factors of this appendix use the truncated expansion without the term of Appendix~\ref{app:norm}, and the distributions of $\mathcal{B}_i$ in Fig.~\ref{fig:bf_orders} use it.

\begin{figure*}[!htbp]
\centering
\includegraphics[width=0.49\linewidth]{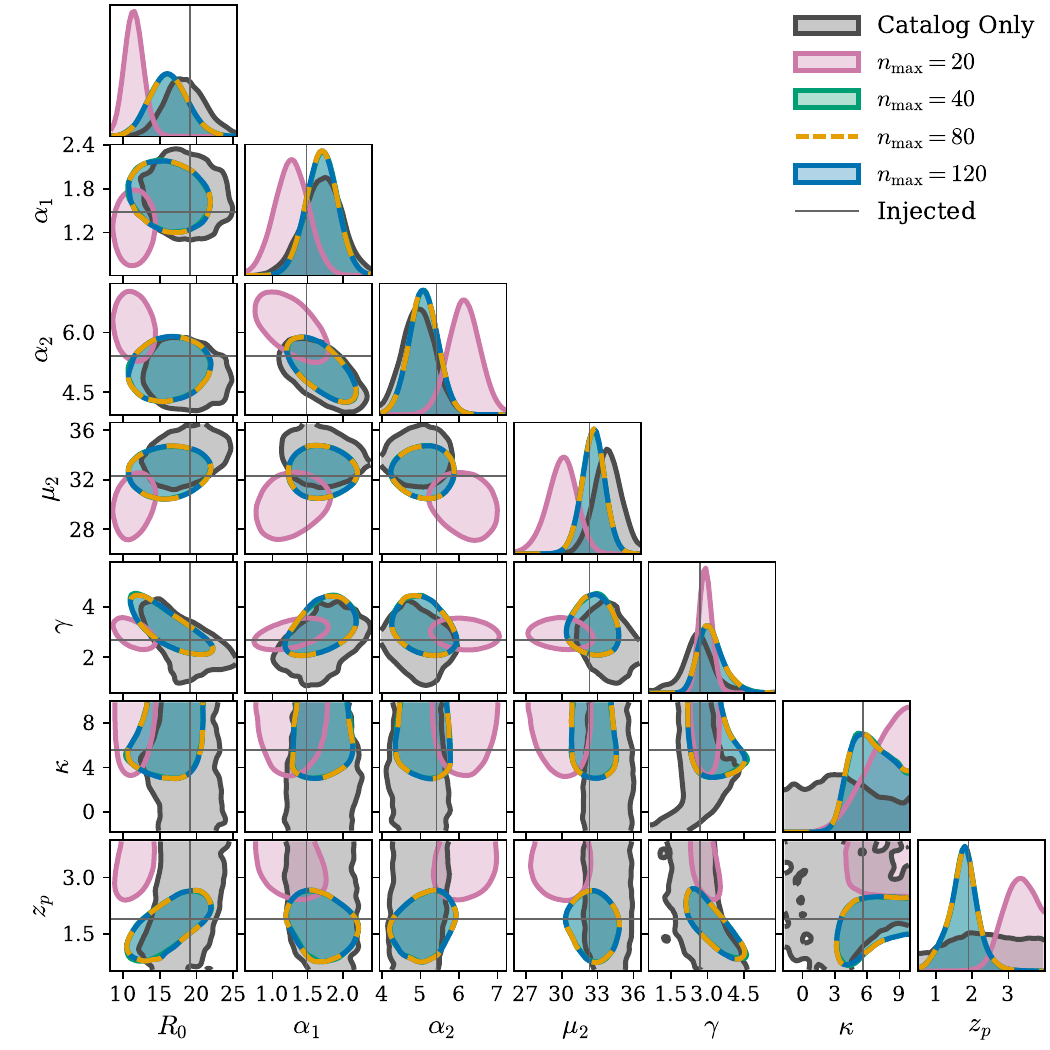}
\includegraphics[width=0.49\linewidth]{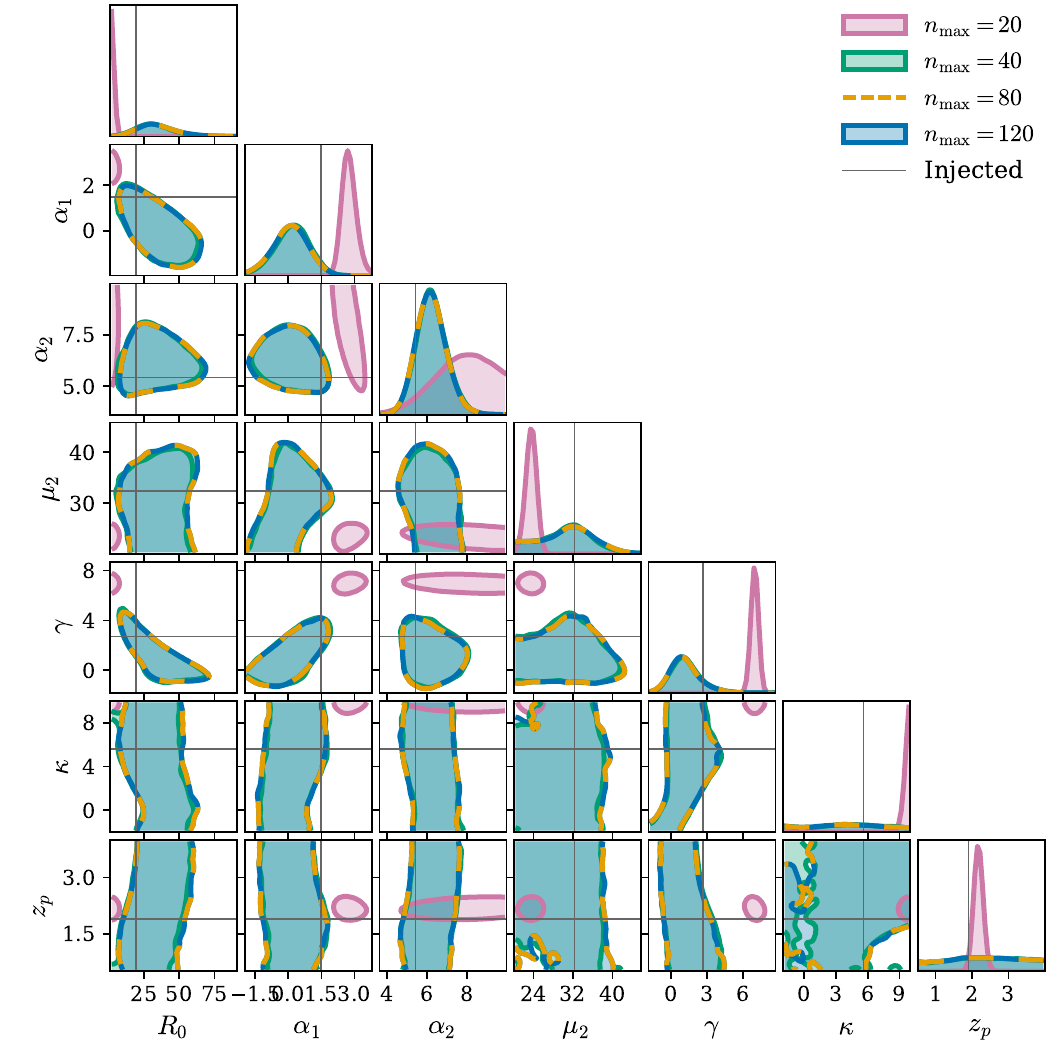}
\caption{Left: Posteriors of the seven population parameters in the catalog-informed search for $\rho_{\rm th} = 8$ and four values of $n_{\max}$, with 90 percent credible regions.
The curve for $n_{\max} = 80$ is dashed on top of that for $n_{\max} = 120$.
The grey regions show the catalog-only posteriors, and the grey lines mark the injected values.
Right: Same as the left panel, but for the uninformed search with $\rho_{\rm th} = 8$.}
\label{fig:pop_inf8}
\end{figure*}

\begin{figure*}[!htbp]
\centering
\includegraphics[width=0.49\linewidth]{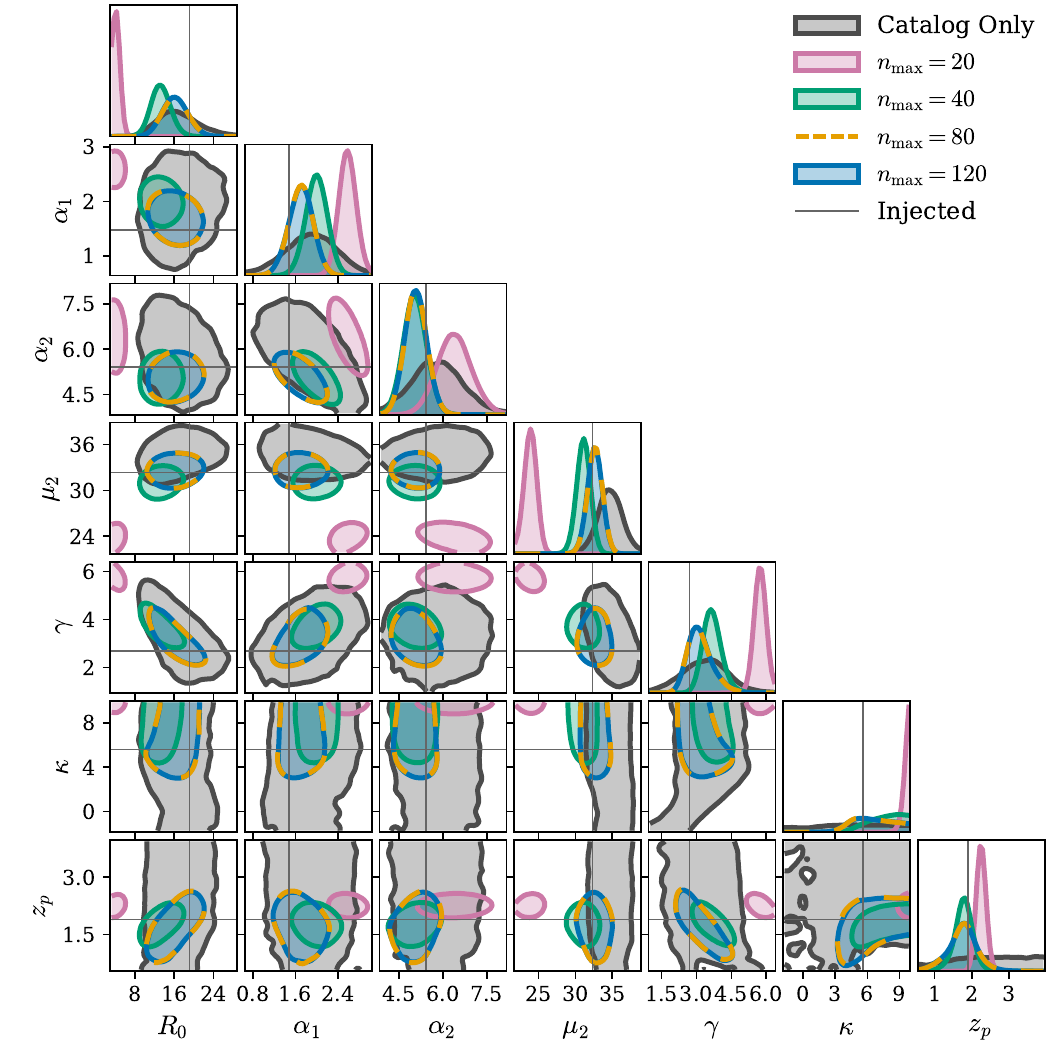}
\includegraphics[width=0.49\linewidth]{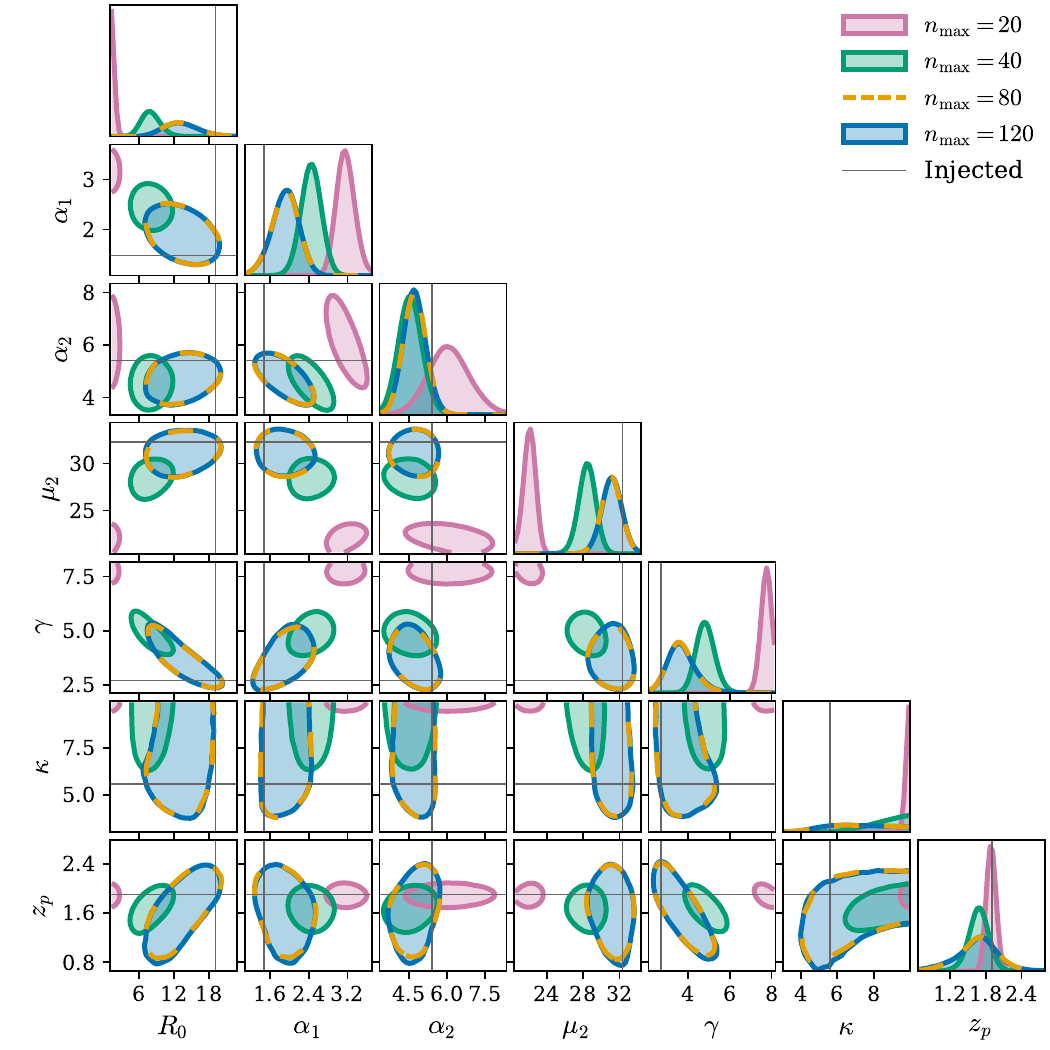}
\caption{Same as Fig.~\ref{fig:pop_inf8}, but for $\rho_{\rm th} = 12$.}
\label{fig:pop_inf12}
\end{figure*}

The two panels of Fig.~\ref{fig:pop_inf8} show the population posteriors of the two searches for $\rho_{\rm th} = 8$ and four values of $n_{\max}$.
The posteriors for $n_{\max} = 40$, 80, and 120 coincide, apart from the unconstrained plane of $\kappa$ and $z_p$ in the uninformed search, and the posterior for $n_{\max} = 20$ is displaced and reaches the edges of the prior in the uninformed search.
Similarly, the two panels of Fig.~\ref{fig:pop_inf12} show the population posteriors of the two searches for $\rho_{\rm th} = 12$, where 127 segments are detected and the posteriors for $n_{\max} = 80$ and 120 coincide, as expected from $n_{\max} > 72$.

\begin{figure}[!htbp]
\centering
\includegraphics[width=0.49\linewidth]{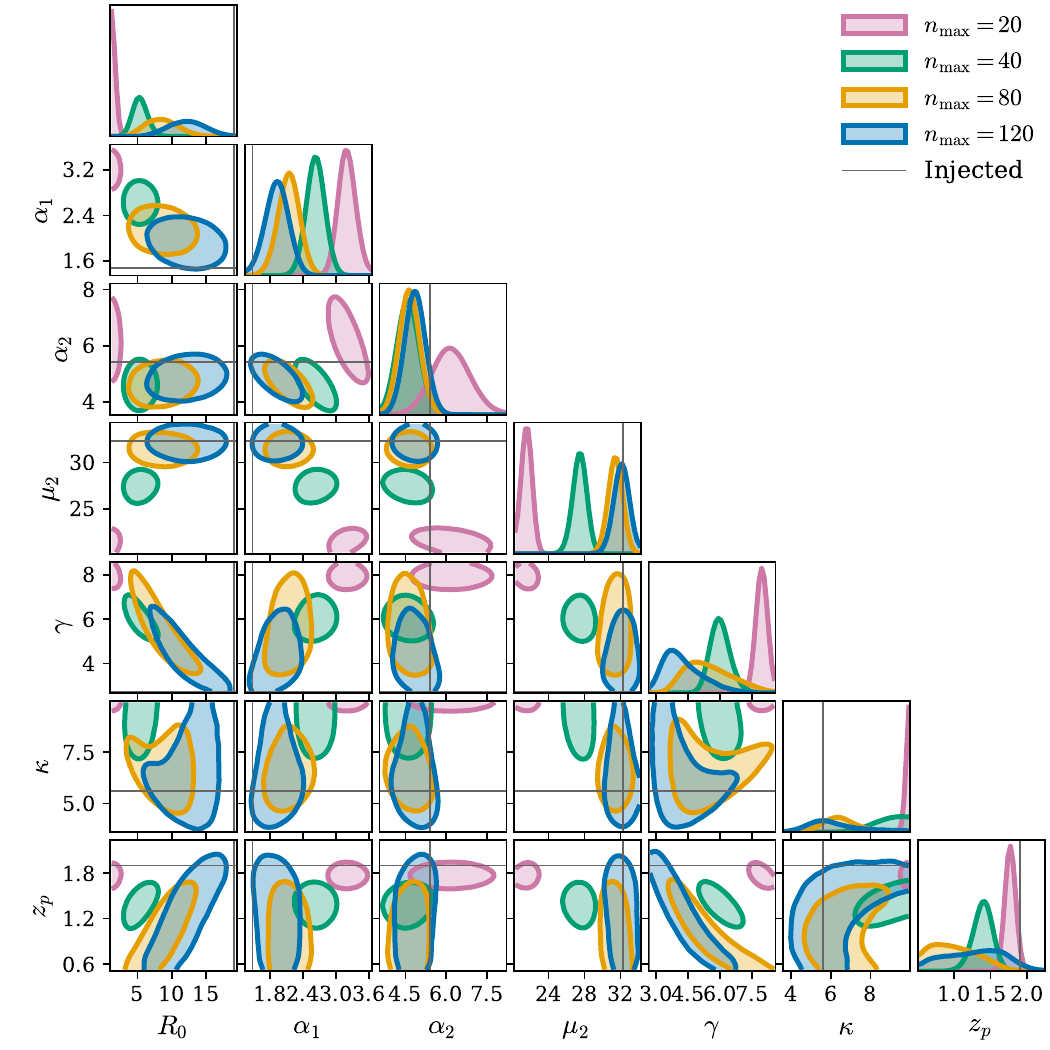}
\caption{Same as Fig.~\ref{fig:pop_inf8}, but for an analysis with no threshold, in which every segment is truncated at $n_{\max}$ and the prior is $\pi_0$.
The catalog-only posteriors are not shown in this plot, and the curve for $n_{\max} = 80$ is solid as opposed to dashed.}
\label{fig:pop_all}
\end{figure}

\begin{figure*}[!htbp]
\centering
\includegraphics[width=\textwidth]{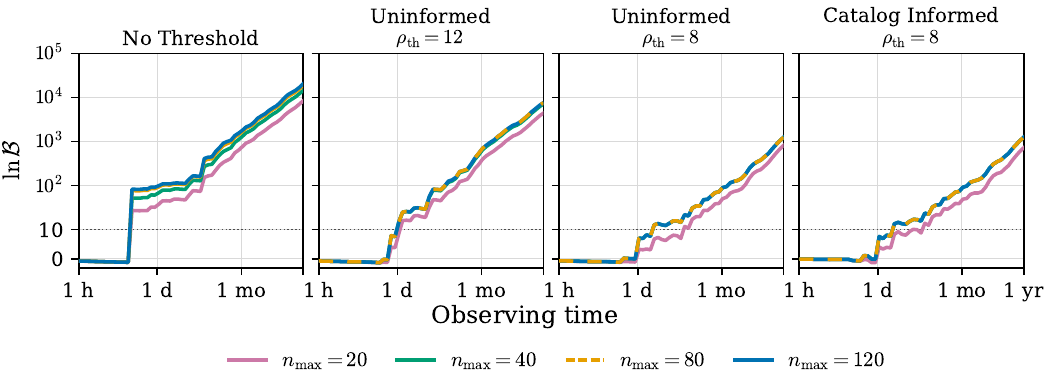}
\caption{Accumulation of the log Bayes factor with observing time for four values of $n_{\max}$ in one year of simulated data.
From left to right, the panels show every segment truncated at $n_{\max}$ with no threshold, the uninformed search $\ln\mathcal{B}_{\rm U}$ for $\rho_{\rm th} = 12$ and 8, and the catalog-informed search $\ln\mathcal{B}_{\rm I}$ for $\rho_{\rm th} = 8$.
The dotted line marks $\ln\mathcal{B} = 10$, and the vertical axis is linear below this line and logarithmic above it.
In the two rightmost panels, the $n_{\rm max}=40$ curve is hidden by the $n_{\rm max}=80$ and 120 curves.}
\label{fig:time5}
\end{figure*}

Figure~\ref{fig:pop_all} shows the analysis with no threshold, where the posteriors of the four orders differ from each other.
For the previous two cases, which had finite threshold values $\rho_{\rm th}=8$ and 12, the orders above $n_{\min}$ had the same posteriors and Bayes factors that agreed within two units of $\ln\mathcal{B}$ in the searches of Fig.~\ref{fig:time5}.
However, for the analysis with no threshold, $n_{\min}$ is set by the loudest merger of the simulation and lies above $n_{\max} = 120$.

\begin{figure*}[!htbp]
\centering
\includegraphics[width=0.5\columnwidth]{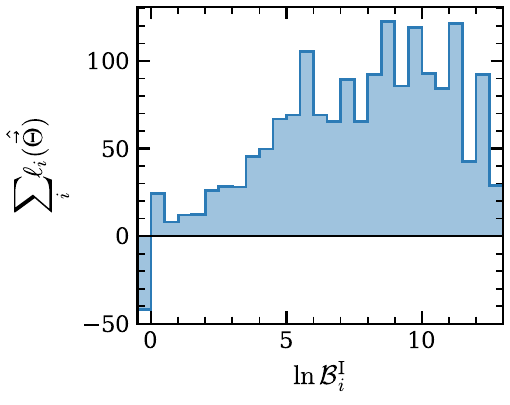}
\caption{Sum of $\ell_i(\hat{\vec\Theta}) = \ln\big[1 + \bar N(\hat{\vec\Theta})\,\big(\bar\Lambda_i(\hat{\vec\Theta}) - 1\big)\big]$ over the unresolved mergers in intervals of $\ln\mathcal{B}_i^{\rm I}$ for $n_{\max} = 120$.}
\label{fig:bf_mix}
\end{figure*}

Figure~\ref{fig:bf_mix} shows how the log likelihood ratio of the unresolved mergers is distributed over $\ln\mathcal{B}_i^{\rm I}$, and how most of the log likelihood ratio comes from the mergers with $\ln\mathcal{B}_i^{\rm I} > 3$. Figure~\ref{fig:bf_orders} shows the distribution of $\mathcal{B}_i$ of Fig.~\ref{fig:year_lnb} at lower orders.
At the first order (i.e., $n_{\max} = 1$ of the cross-correlation), no segment exceeds $\ln\mathcal{B}_i = 0.03$ and the first order cannot tell a single merger from noise.
Already at $n_{\max} = 10$ there are $\mathcal{O}(100)$ unresolved mergers above the largest value of the noise-only segments, and the tail grows with the order until it stops changing above $n_{\min}$.

Figure~\ref{fig:time5} extends
Fig.~\ref{fig:time_th} to include an analysis with no threshold as well as four different values of $n_{\rm max}$.
The uninformed search for $\rho_{\rm th} = 12$ reaches $\ln\mathcal{B}_{\rm U} = 10$ after about one day and ends the year at $\ln\mathcal{B}_{\rm U} \simeq 7.6 \times 10^3$.
The curves of the orders with $n_{\max} > \rho_{\rm th}^2/2$ lie on top of each other for both thresholds.
The curve for $n_{\max} = 20$ rises more slowly in every panel, and for $\rho_{\rm th} = 12$ the curve for $n_{\max} = 40$ ends the year 7 percent below those above $n_{\min}$.
With no threshold, the segments above $\rho_{\rm mf}^{\max} = \sqrt{2 n_{\max}}$ are truncated as well, and the Bayes factor still grows from $n_{\max} = 80$ to 120.

\begin{figure*}[!htbp]
\centering
\includegraphics[width=\textwidth]{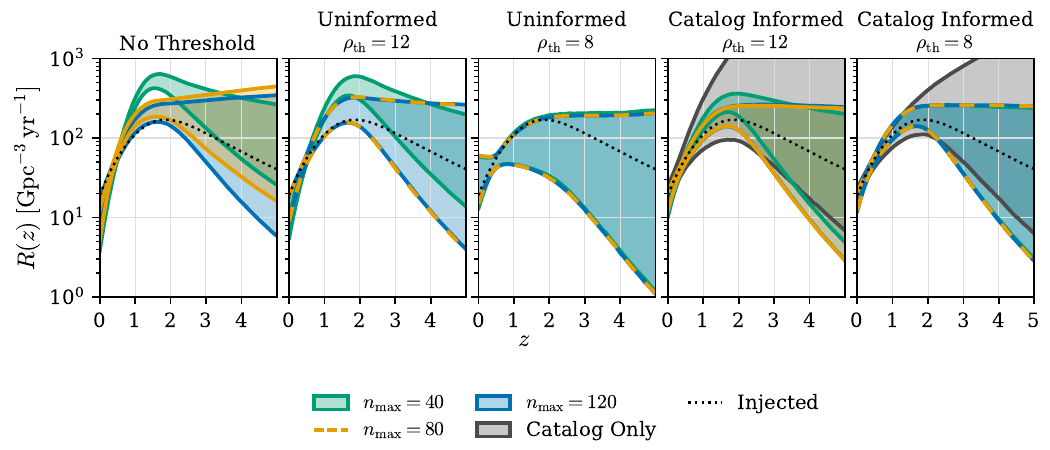}
\caption{The merger rate density $R(z)$ of~\eqref{eq:rz} over the posteriors of the five cases for three values of $n_{\max}$, as bands between the 5th and the 95th percentile at every redshift.
From left to right, the panels show every segment truncated at $n_{\max}$ with no threshold, the uninformed search for $\rho_{\rm th} = 12$ and 8, and the catalog-informed search for $\rho_{\rm th} = 12$ and 8.
The grey bands show the catalog-only posteriors, and the dotted curve is the injected rate history.
For the two panels having $\rho_{\rm th}=8$, the $n_{\rm max}=40$ bands lie behind the $n_{\rm max}=80$ and 120 bands.}
\label{fig:rz5}
\end{figure*}

Figure~\ref{fig:rz5} shows the merger rate density $R(z) = R_0\, \psi(z \mid \vec\theta)$ of~\eqref{eq:rz} over the posteriors of the two searches for both thresholds and of the analysis with no threshold, for $n_{\max} = 40$, 80, and 120.
For the catalog-informed searches (the two right panels), the band of the catalog-only analysis has no upper bound within the plotted range at high redshift.
For the uninformed searches, which analyze only the unresolved segments, the rate density stays below about $300~{\rm Gpc^{-3}\,yr^{-1}}$ at every redshift for the orders above $n_{\min}$.

\section{Noise-only tests}
\label{app:null}

To check that the search does not claim a background under the null hypothesis, we rerun our analysis on ten years of Gaussian noise alone with the same noise spectrum of Sec.~\ref{sec:population}.
We found no segment of any of these years exceeding $\rho_{\rm th} = 8$, so the catalog is empty and the catalog-informed search reduces to the analysis with no threshold of Appendix~\ref{app:other-cases}.
We also run on these ten years the uninformed search that removes the segments above the threshold and conditions the others on staying below it as in~\eqref{eq:LU}.

\begin{figure*}[!tp]
\centering
\includegraphics[width=0.8\textwidth]{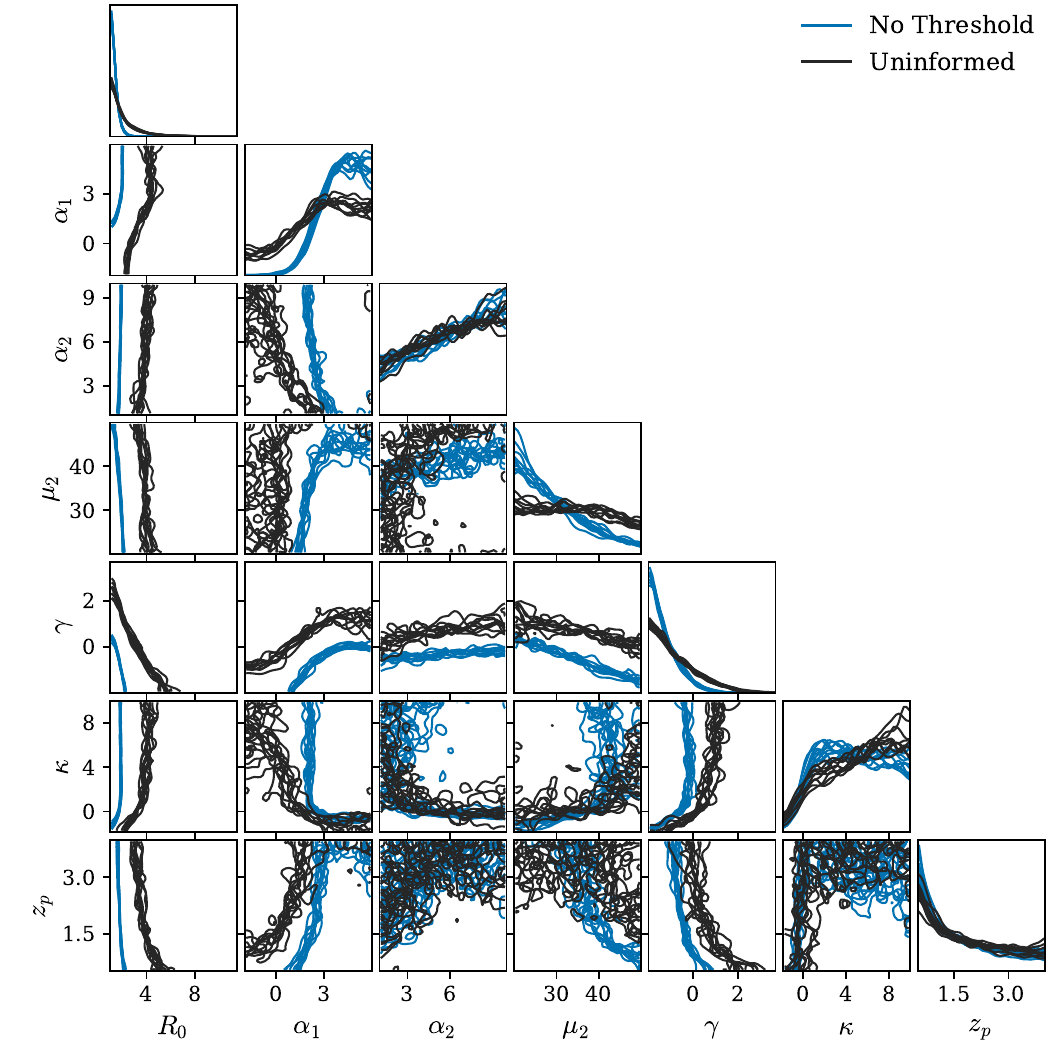}
\caption{Posteriors of the population parameters for ten years of Gaussian noise.
Plotted are the 90 percent credible regions of every year.
The blue regions show the analysis with no threshold, which equals the catalog-informed search when no segment is detected.
The black regions show the results of the analysis of the uninformed search.}
\label{fig:null}
\end{figure*}

Figure~\ref{fig:null} shows the posteriors of the seven population parameters of the two analyses in every year.
In both analyses, the rate density $R_0$ rests against the lower edge $1~{\rm Gpc^{-3}\,yr^{-1}}$ of its prior, and the posteriors of the ten years nearly coincide.
In every year, the analysis with no threshold gives a log Bayes factor between $-9.5$ and $-9.3$ against noise only and bounds the rate at $R_0 < 1.7~{\rm Gpc^{-3}\,yr^{-1}}$ (90 percent upper limit) far below the injected value $19.1~{\rm Gpc^{-3}\,yr^{-1}}$ of Sec.~\ref{sec:population}.
The uninformed search gives a wider posterior with $R_0 < 4.1~{\rm Gpc^{-3}\,yr^{-1}}$ and a log Bayes factor between $-3.6$ and $-3.4$, because conditioning on the absence of detections removes the information that no segment was loud.
Since the prior does not allow a vanishing rate, the six shape parameters move toward the populations that produce the fewest detectable mergers at a given rate.
These populations have a large low-mass slope $\alpha_1$ and a low second peak $\mu_2$, and their rate rises slowly with redshift and peaks early.

\bibliography{refs}

\end{document}